%% file: Agentification.tex
\documentclass[11pt]{article}
\usepackage[margin=1in]{geometry}
\usepackage[T1]{fontenc}
\usepackage[utf8]{inputenc}
\usepackage{lmodern}
\pdfmapfile{+lm.map}
\pdfmapfile{+amsfonts.map}
\usepackage{amsmath,amssymb,amsthm,booktabs,array,longtable}
\usepackage[round,authoryear]{natbib}
\usepackage[expansion=false]{microtype}
\usepackage[hidelinks]{hyperref}
\usepackage{setspace}
\usepackage{enumitem}
\usepackage{graphicx}
\usepackage[section]{placeins}
\setlist{nosep}
\newtheorem{proposition}{Proposition}
\newcommand{\dd}{\mathrm d}
\newcommand{\R}{\mathcal R}
\newcommand{\PC}{\mathcal P}
\hypersetup{pdftitle={Burning the Bridge: Agentification and Domestic Value Added},pdfauthor={Eduardo Levy Yeyati},pdfsubject={Imported AI tasks, domestic value added, and foreign dependence}}
\title{\textbf{Burning the Bridge}\\[0.4em]\large Agentification and Domestic Value Added}
\author{Eduardo Levy Yeyati\\\small Universidad Torcuato Di Tella and Inter-American Development Bank}
\date{This version: November 2026}
\begin{document}
\maketitle
\begin{abstract}
Firms that replace local work with foreign AI agents can produce more per paid hour while generating less income at home. We study this trade-off in a small open economy where exporters and local services both use agents. Once wages reach a reservation floor, extending agent use in local services lowers domestic income at unchanged foreign costs. Value added at constant prices and consumption also fall if suppliers capture enough of the saving. Exporters have an offset: adopting agents can expand production and earn the foreign exchange needed to pay for them. The difference is substantial. Using Brazilian accounts as a reference, a supplier bargaining share above about 0.15 is sufficient to make further local-service adoption reduce value added even at negligible foreign costs. The corresponding threshold in tradables is about 0.66. We derive separate export-growth requirements for preserving income and consumption and illustrate them using historical accounts for Brazil and Mexico. Cheaper agents ease the consumption requirement; lower export prices raise it. Over time, declining domestic capabilities can also make returning to local production more expensive, allowing suppliers to charge more at renewal.
\end{abstract}
\newpage

\input{introduction}

\section{An economy that imports agents}
\label{sec:model}
The gain from cheaper agents depends on how production expands and how the labor market absorbs displacement.

\subsection{Goods, households, and the reservation wage}
There are three final goods. Good $T$ is tradable at world price $p_T=1$. Good $N$ is a local service produced with tasks that agents may execute. Good $R$ is a residual local activity requiring human labor. It represents personal services or care and can absorb displaced workers as long as domestic demand sustains their earnings. Prices $p_N$ and $p_R$ clear domestic markets.

Residents own the fixed domestic factors. Workers in the modeled segment have homogeneous labor capacity with potential hours $\bar L$. These hours need not equal the entire working-age population's potential hours. Hours are divided between paid work and nonemployment. For a fixed resident population, public support combines universal grants with an hours-contingent supplement. Total outlay is $B_U+b(\bar L-L)$, where $B_U\ge0$ is the aggregate universal component and $b\ge0$ is the payment per unused capacity-hour. The universal grant is retained when an hour is worked; the supplement falls by $b$. Coverage is complete within the modeled population.

All residents have the same within-period consumption aggregator,
\begin{equation}
 C=\prod_{j\in\{T,N,R\}} C_j^{\gamma_j},
 \qquad \gamma_j>0,\quad \sum_j\gamma_j=1.
 \label{eq:preferences}
\end{equation}
Let $r>0$ be the net earnings threshold for supplying a paid hour, with $0\le b\le r$. The reservation threshold and the two fiscal parameters are separate primitives. The benchmark holds $r$ fixed when $B_U$ changes, excluding income effects of transfers on participation. Let $\tau_L\in[0,1)$ tax gross labor earnings and $\tau_V\ge0$ tax final consumption uniformly. The labor-market closure is
\begin{equation}
 \underline w=\frac{r}{1-\tau_L},\qquad
 w\ge\underline w,\quad L\le\bar L,\quad
 (w-\underline w)(\bar L-L)=0.
 \label{eq:floor}
\end{equation}
At wages above the threshold, all potential hours of this segment are employed. At the threshold, labor demand can leave unused hours. The latter state can include jobseekers and residents who do not search. Aggregate unemployment and inactivity statistics alone do not identify this branch, because their other causes are outside this homogeneous-segment model.

The benchmark fixes $r$, $B_U$, and $b$ in monetary units of the tradable numeraire. In consumption units, the reservation comparison divides both $(1-\tau_L)w$ and $r$ by the same consumer-price index. A uniform VAT therefore cancels from that comparison. The nonfiscal component $r-b$ is an earnings-equivalent reservation term, excluded from measured value added and market consumption; it never enters aggregate spending as a resource. Public transfers are financed explicitly. The rest of the reservation threshold is a reduced-form labor-supply restriction, without a complete model of household nonemployment budgets or welfare.

A fixed physical quantity of home production retains its direct usefulness when VAT rises, and purchasing a fixed quantity of replacement care generally becomes more expensive. Either mechanism can make participation respond to VAT. A fixed monetary expenditure can share the earnings deflator; a fixed quantity of purchased care does not establish that neutrality. The benchmark takes the nominal threshold as given. Indexing the reservation threshold or support parameters to consumer prices would require solving fiscal policy and participation jointly.

\subsection{Task execution and domestic production}
For $j\in\{T,N\}$, competitive firms produce using a fixed factor $K_j>0$ and a unit interval of tasks:
\begin{equation}
 Y_j=A_jK_j^{\alpha_j}
       \exp\left\{\beta_j\int_0^1\log x_j(i)\,\dd i\right\},
 \qquad \beta_j=1-\alpha_j,\quad 0<\alpha_j<1.
 \label{eq:production}
\end{equation}
Human execution supplies $x_j(i)=h_j l_j(i)$, with $h_j>0$ given in this section. An eligible task can instead be supplied by an imported agent, $x_j(i)=q_j(i)$. Quantities and prices are measured per effective execution unit. Task eligibility has measure $\eta_j\in(0,1)$, and the actual agent-executed measure is $m_j\le\eta_j$. Domestic labor moves freely between tasks and sectors at wage $w$.

Let $P_j$ be the net foreign price and $t\ge0$ the fixed ad valorem tax on agent purchases. Human and agent execution costs per unit are
\begin{equation}
 C_j^H=\frac{w}{h_j},\qquad C_j^A=(1+t)P_j.
 \label{eq:costs}
\end{equation}
Foreign marginal cost is $c_{Aj}>0$. A unit procurement contract divides the saving over the human fallback according to
\begin{equation}
 P_j=\frac{\zeta_j}{1+t}\frac{w}{h_j}+(1-\zeta_j)c_{Aj},
 \qquad 0<\zeta_j<1.
 \label{eq:bargain}
\end{equation}
Here $\zeta_j$ is the supplier's bargaining weight. This rule follows from generalized Nash bargaining over the saving on one required execution unit, as shown in Appendix \ref{app:bargain}. The quantity of execution is subsequently chosen by the firm at the quoted unit price. 

The results use the strict cost-advantage regime
\begin{equation}
 (1+t)c_{Aj}<\frac{\underline w}{h_j}.
 \label{eq:eligible}
\end{equation}
It implies $C_j^A<C_j^H$ at all equilibrium wages, so eligible tasks are voluntarily agentified: $m_j=\eta_j$. Firms choose the cheaper execution method, and technology determines which tasks agents can perform. Progress expands $\eta_T$ and $\eta_N$ and can lower both foreign marginal costs. The set of human tasks is fixed in the benchmark. The competition limit is $\zeta_j=0$; the interior rule retains a positive provider margin without requiring capability erosion. With zero access charges and human-equivalent execution units, $\lim_{\zeta_j\to1}(1+t)P_j=w/h_j$ recovers the binding price-of-indifference boundary in \citet{AtheyScottMorton2025}. At that limit the strict procurement saving vanishes. This price condition is shared across the two frameworks; the production and entry structures differ.

Cost minimization and the fixed-factor optimum imply, writing $\R_j=p_jY_j$,
\begin{align}
 G_j&=(w/h_j)^{1-m_j}[(1+t)P_j]^{m_j},
 \label{eq:G}\\
 Y_j&=K_jA_j^{1/\alpha_j}
           (\beta_jp_j/G_j)^{\beta_j/\alpha_j},
 \label{eq:supply}\\
 wL_j&=\ell_j\R_j,\quad
 (1+t)P_jq_j=\beta_jm_j\R_j,\quad
 \Pi_j=\alpha_j\R_j,
 \qquad \ell_j=\beta_j(1-m_j).
 \label{eq:shares}
\end{align}
Each task receives expenditure $\beta_j\R_j$ per unit of task measure. The gross execution share transferred out of domestic labor by an increase in $m_j$ is therefore $\beta_j\,\dd m_j$. Only $1/(1+t)$ of the corresponding agent payment goes abroad.

The residual activity has no agent-eligible tasks:
\begin{equation}
 Y_R=A_RK_R^{\alpha_R}L_R^{\beta_R},\qquad
 wL_R=\beta_Rp_RY_R,\quad \Pi_R=\alpha_Rp_RY_R.
 \label{eq:residual}
\end{equation}
Its labor marginal product falls at a given output price. Its equilibrium price also responds to domestic demand. The model retains both margins.

\subsection{National accounts, foreign exchange, and the fiscal budget}
Foreign-agent payments and their sectoral revenue shares are
\begin{equation}
 I=\sum_{j=T,N}P_jq_j,\qquad
 \mu_j=\frac{P_jq_j}{\R_j}=\frac{\beta_jm_j}{1+t}.
 \label{eq:imports}
\end{equation}
Define domestic factor income and national value added excluding final consumption tax as
\begin{equation}
 Y^D=wL+\Pi,\qquad
 V=\sum_jp_jY_j-I=Y^D+tI,
 \qquad \Pi=\sum_j\Pi_j.
 \label{eq:VA}
\end{equation}
$Y^D$ is value added at factor cost. $V$ includes agent-tax revenue and excludes final VAT. At $t=0$ the two coincide. Holding tax rates fixed avoids attributing a change in production to a change in tax valuation.

Let $F$ denote net external financing, positive for a resource inflow. The goods and external accounts are
\begin{equation}
 C_N=Y_N,\quad C_R=Y_R,\quad
 C_T+I=Y_T+F.
 \label{eq:external}
\end{equation}
The benchmark fixes $F=0$, with the floor financed from domestic resources. A fixed nonzero $F$ is introduced only for the explicit counterfactual in Section \ref{sec:external_result}.

VAT applies uniformly to final domestic consumption, including consumption financed by benefits. Agent purchases receive intermediate-input credits; exports are zero rated. There are no other government purchases. Let $\mathcal S\ge0$ be a surplus rebate to fixed-factor owners. The fiscal budget is
\begin{equation}
 B_U+b(\bar L-L)+\mathcal S
   =\tau_LwL+tI+\tau_V E,
 \qquad E=\sum_jp_jC_j=V+F.
 \label{eq:fiscalbudget}
\end{equation}
The rebate prevents a negative required VAT from becoming an unannounced consumption subsidy. Define
\begin{equation}
 \tau_V^{\rm req}
 =\frac{B_U+b(\bar L-L)-\tau_LwL-tI}{E},\qquad
 \tau_V=\max\{0,\tau_V^{\rm req}\},\quad
 \mathcal S=E\max\{0,-\tau_V^{\rm req}\}.
 \label{eq:reqVAT}
\end{equation}
A specified upper bound $\bar\tau_V$ restricts the analysis to equilibria satisfying $\tau_V^{\rm req}\le\bar\tau_V$. The maintained floor regime ends at this feasibility boundary.

Under \eqref{eq:preferences}--\eqref{eq:reqVAT}, aggregate producer-price demand satisfies
\begin{equation}
 p_jC_j=\gamma_jE,\qquad E=V+F.
 \label{eq:demand}
\end{equation}
Conditional on feasibility, the real allocation can be solved independently of $\tau_V$ and $\mathcal S$. This result holds for a floor fixed in the tradable numeraire, common expenditure weights, and a uniform fully credited consumption tax.

The neutrality is an aggregation result. The tax changes individual purchasing power and redistributes consumption between owners, workers, and beneficiaries. It also lowers the real purchasing power of the monetary reservation threshold and of the public benefit. Different expenditure shares, reservation or benefit indexation, a nonmarket-production outside option, incomplete coverage, or nonuniform taxes would generally feed fiscal policy back into production.

An equilibrium consists of task sourcing, quantities, a common wage, local prices, consumption, and fiscal transfers satisfying firm optimization, \eqref{eq:floor}, goods and external clearing, and the budget. The strict regime \eqref{eq:eligible} fixes $m_j=\eta_j$ through voluntary sourcing. Conditional equilibria can therefore be compared as the technological frontier and foreign costs change.

\section{Results}
\label{sec:results}
The export expansion needed to pay for foreign agents depends on wages and domestic spending.

\subsection{The aggregate multiplier and the two labor regimes}
Write
\begin{align}
 \Xi&=\gamma_T+\gamma_N\mu_N,\qquad
 k=\frac{1-\mu_T}{\Xi},
 \label{eq:Xi}\\
 s_L&=\frac{\ell_T}{1-\mu_T}\Xi
            +\ell_N\gamma_N+\beta_R\gamma_R,
 \label{eq:sl}\\
 s_K&=\frac{\alpha_T}{1-\mu_T}\Xi
            +\alpha_N\gamma_N+\alpha_R\gamma_R.
 \label{eq:sk}
\end{align}
Here $s_L=wL/E$ and $s_K=\Pi/E$ in the balanced-trade benchmark. At $t>0$ these are shares of $V$, which includes agent-tax revenue. They satisfy $s_L+s_K+tI/E=1$.

\begin{proposition}[Allocation, reallocation, and the reservation wage]
\label{prop:allocation}
With $F=0$ and \eqref{eq:eligible}, the conditional real equilibrium is unique and satisfies
\begin{equation}
 E=kY_T(w),\qquad
 L^d(w)=\frac{s_LkY_T(w)}{w},\qquad
 \R_N=\gamma_NE,\quad\R_R=\gamma_RE.
 \label{eq:allocation}
\end{equation}
If $L^d(\underline w)\le\bar L$, the equilibrium has $w=\underline w$ and $L=L^d(\underline w)$. Otherwise the wage is the unique solution of $L^d(w)=\bar L$ above $\underline w$. Fiscal feasibility is checked afterward using \eqref{eq:reqVAT}.

At fixed $t$, an increase in either task share lowers the wage bill as a share of national resources:
\begin{align}
 \frac{\partial s_L}{\partial m_T}
   &=-\frac{\beta_T(\alpha_T+t)\Xi}
          {(1+t)(1-\mu_T)^2}<0,\qquad
 \frac{\partial s_L}{\partial m_N}
   =-\frac{\beta_N\gamma_N(\alpha_T+t)}
          {(1+t)(1-\mu_T)}<0.
 \label{eq:sl_derivatives}
\end{align}
Along the full-employment branch, residual hours equal
\begin{equation}
 L_R=\frac{\beta_R\gamma_R}{s_L}\bar L,
 \label{eq:residual_absorb}
\end{equation}
and rise with agentification. Between two full-employment equilibria, the wage falls exactly when
\begin{equation}
 \frac{E_1}{E_0}<\frac{s_{L0}}{s_{L1}}.
 \label{eq:wagefall}
\end{equation}
On the floor branch, $L=s_LE/\underline w$. Consequently, if task shares weakly rise and $E$ falls, aggregate paid hours and residual hours fall.

\end{proposition}

The multiplier incorporates all three local activities. Human-only residual production leaves its expenditure weight outside the import-leakage denominator $\Xi$. It affects labor demand, rents, and quantities through $s_L$, $s_K$, and $\gamma_R$. Domestic demand is paid for by resident resources; the model supplies no independent expenditure expansion after domestic labor income has been displaced.

A technology path that lowers the full-employment wage can first reallocate hours toward the residual activity and then reduce total paid hours once the floor binds. A sufficiently strong expansion of national income can raise wages or keep the floor inactive. The ordering depends on the technology path.

\input{joint_results}

\subsection{Finite comparisons, the tradable offset, and paid-hour productivity}
\label{sec:finite}
Consider two equilibria at the wage floor, with unchanged $A_j,K_j,h_j,r,B_U,b,\tau_L,t$ and $F=0$. Technology can expand both task frontiers and lower both foreign marginal costs. At $w=\underline w$, \eqref{eq:bargain} fixes the quote for each foreign cost; quotes can differ between the two equilibria when foreign costs change. Let
\begin{equation}
 \mathcal E=\frac{E_1}{E_0},\qquad
 g_j=\frac{G_{j1}}{G_{j0}},\qquad
 r_j=\frac{Y_{j1}}{Y_{j0}}.
 \label{eq:ratios}
\end{equation}
Then the exact finite changes are
\begin{align}
 r_T&=g_T^{-\beta_T/\alpha_T},
 &\mathcal E&=r_T\frac{1-\mu_{T1}}{1-\mu_{T0}}\frac{\Xi_0}{\Xi_1},
 \label{eq:finiteE}\\
 r_N&=(\mathcal E/g_N)^{\beta_N},
 &r_R&=\mathcal E^{\beta_R}.
 \label{eq:finiteY}
\end{align}

\begin{proposition}[The tradable offset and paid-hour productivity]
\label{prop:threshold}
National income falls if and only if
\begin{equation}
 r_T<\frac{1-\mu_{T0}}{1-\mu_{T1}}\frac{\Xi_1}{\Xi_0}.
 \label{eq:Tthreshold}
\end{equation}
Reversing this inequality preserves or raises income. If $m_{j1}\ge m_{j0}$, $g_N\le1$, and $\mathcal E<1$, aggregate paid hours fall and output per paid hour rises in $N$. It also rises in $T$ whenever $m_T$ rises strictly. Specifically,
\begin{align}
 \frac{(Y_T/L_T)_1}{(Y_T/L_T)_0}
   &=\frac{1-m_{T0}}{1-m_{T1}},
 \label{eq:prodT}\\
 \frac{(Y_N/L_N)_1}{(Y_N/L_N)_0}
   &=\mathcal E^{-\alpha_N}g_N^{-\beta_N}
                \frac{1-m_{N0}}{1-m_{N1}}>1.
 \label{eq:prodN}
\end{align}
Nontradable task output itself rises if and only if $g_N<\mathcal E$; paid-hour productivity can rise even when gross output falls.

Define $\mathcal H_E=(1-\mu_{T0})(\Xi_1/\Xi_0)/(1-\mu_{T1})$. The exact consumption comparison on the same floor branch is
\begin{equation}
 \frac{Q_1}{Q_0}=\left(\frac{r_T}{\mathcal H_E}\right)^B
                  g_N^{-\beta_N\gamma_N},\qquad
 \mathcal H_Q=\mathcal H_E g_N^{\beta_N\gamma_N/B}.
 \label{eq:consumption_hurdle}
\end{equation}
Real consumption is preserved or increased if and only if $r_T\ge\mathcal H_Q$. When local execution becomes cheaper, $g_N<1$, this hurdle is lower than the income-preserving hurdle $\mathcal H_E$.
\end{proposition}

The tradable response in \eqref{eq:Tthreshold} is endogenous. A lower $\alpha_T$ makes tradable supply more responsive to lower execution costs. Concentrating progress in $T$ improves income when the resulting expansion meets \eqref{eq:Tthreshold}. Small $\gamma_N\mu_N$ reduces the leakage attached to local service demand.

Equation \eqref{eq:consumption_hurdle} distinguishes the export requirement from the export response. The requirement combines the imported content of demand with the consumption gain from cheaper local execution; the response follows from tradable supply. A country can fail the income requirement while meeting the consumption requirement. Both comparisons are needed when imported agents become cheaper.

The exact volume comparison is
\begin{equation}
\begin{split}
 V^Q_1-V^Q_0={}&Y_{T0}(r_T-1)
  +\gamma_NE_0(r_N-1)+\gamma_RE_0(r_R-1)\\
 &-\sum_{j=T,N}P_{j0}(q_{j1}-q_{j0}).
\end{split}
 \label{eq:finiteVQ}
\end{equation}
Equations \eqref{eq:finiteE}--\eqref{eq:finiteVQ} evaluate current-price income and reference-price production separately in finite comparisons.

For a fixed positive agent tax, the production envelope uses the gross execution gap $\delta_j=\log[(\underline w/h_j)/((1+t)P_j)]$ and gives
\begin{equation}
 \dd V^Q=\underline w\,\dd L+t\sum_{j=T,N}P_j\,\dd q_j
       +\sum_{j=T,N}\beta_j\R_j\delta_j\,\dd m_j.
 \label{eq:volume_envelope}
\end{equation}
The last term is the resource saving from tasks newly opened to agents. Appendix \ref{app:finitebounds} derives exact bargaining bounds for finite comparisons at initial reference prices. Those conditions complement the local thresholds in Proposition \ref{prop:joint}.

\subsection{Foreign-exchange resources and financing}
\label{sec:external_result}
With $F=0$, net exports of the tradable good are
\begin{equation}
 X=Y_T-\gamma_TE=I.
 \label{eq:netexports}
\end{equation}
If $Y_T$ rises and $E$ falls, agent imports and net exports rise, while domestic tradable consumption falls. Balanced trade turns the higher import bill into a reallocation toward net exports and away from domestic tradable absorption.

There are two distinct counterfactual financing calculations. At the new production allocation, preserving the old tradable consumption requires the arithmetic resource gap $\Delta I-\Delta Y_T=-\gamma_T\Delta E>0$. If local demand is also allowed to respond to the financing, \eqref{eq:allocation} generalizes to
\begin{equation}
 E=\frac{(1-\mu_T)Y_T+F}{\Xi}.
 \label{eq:withF}
\end{equation}
Provided the floor remains active and the prior capability stock is fixed, the actual financing needed to restore $E_0$ at the new technology is therefore
\begin{equation}
 F^*=\Xi_1(E_0-E_1)>0.
 \label{eq:Fstar}
\end{equation}
It includes the additional agent imports induced by restoring local expenditure. The two calculations refer to different allocations. The benchmark uses neither inflow: it keeps $F=0$.

On the adverse branch with fixed human capability, both local prices fall: $p_R$ has log change $\alpha_R\dd\log E$, and
\begin{equation}
 \dd\log p_N=\alpha_N\dd\log E+\beta_N\dd\log G_N.
 \label{eq:localprices}
\end{equation}
Domestic relative prices adjust to the change in demand. Favorable changes in imported-agent prices remain a separate terms-of-trade benefit. The benchmark holds the world price of tradable output fixed.

\input{export_prices}

\paragraph{Public support.}
Universal support becomes harder to finance when domestic income falls, while hours-contingent support also expands as paid hours contract; Appendix \ref{app:fiscal_details} derives both channels under the rule $B_U+b(\bar L-L)$.

\section{Capability, renewal, and external dependence}
\label{sec:dependence}
Agents can become harder to drop than to adopt.

\citet{AtheyScottMorton2025} identify the loss of domestic productive capabilities as an open question in their policy research agenda. This extension makes the capability state depend on equilibrium human execution and links it to the renewal price. \citet{SpencerRaubitschek1996} show how domestic input production can discipline foreign oligopolistic suppliers. \citet{StenbackaTombak2012} study the balance between outsourcing savings and supplier bargaining power.

\subsection{A two-date extension with anticipated renewal}
The one-date model already contains supplier market power. To obtain collective hold-up, introduce dates $s=0,1$ and a shared capability stock $H_{js}$. Set $h_{js}=h_j(H_{js})$, with $h_j'>0$, and let
\begin{equation}
 H_{j1}=(1-\delta_{Hj})H_{j0}+\chi_j L_{j0},
 \qquad 0\le\delta_{Hj}\le1,\quad \chi_j>0.
 \label{eq:Hlaw}
\end{equation}
$L_{j0}$ is domestic task execution in that sector, excluding residual activity. This use-dependent capability technology maps equilibrium domestic execution into a shared future state. Its formation parameters summarize apprenticeship and organizational learning.

For a comparison at fixed initial capability, $\Delta H_{j1}=\chi_j\Delta L_{j0}$. Agentification therefore erodes future capability if it reduces relevant human execution. A sufficient scale expansion can preserve or increase that execution. The capability sign follows the execution-hours response.

Firms correctly anticipate date-1 prices and technology. Each is atomistic relative to the shared capability pool and takes its evolution as given. With no firm-specific installation investment in this extension, the firm's date-0 sourcing decision has no private effect on date-1 capability. Any positive discounting of operating profits consequently preserves the within-date cost-minimizing sourcing choices. Under the strict advantage conditions at both dates, $m_{js}=\eta_{js}$. Correctly anticipated dependence is consistent with decentralized adoption.

The sequence is recursive. Date 0 solves the production equilibrium given $H_0$. Equation \eqref{eq:Hlaw} determines $H_1$. Date 1 solves the same equilibrium with the new human capability, foreign costs, and task frontier. At renewal, the foreign price is
\begin{equation}
 P_{j1}=\frac{\zeta_j}{1+t}\frac{w_1}{h_j(H_{j1})}
                 +(1-\zeta_j)c_{Aj1}.
 \label{eq:renewal}
\end{equation}
If agents remain strictly cheaper than human execution, the wage is still determined by a decreasing labor-demand function. At the floor, renewal prices are explicit. Above the floor, prices and wages are solved jointly through the one-dimensional labor-market equation.

\begin{proposition}[The effective fallback and renewal exposure]
\label{prop:renewal}
Under the unit bargaining rule with fixed $t$ and $\zeta_j$, the effective domestic outside cost rises precisely when
\begin{equation}
 \Delta\log w>\Delta\log h_j(H_j).
 \label{eq:fallback_condition}
\end{equation}
With the wage fixed at the floor, a loss of capability raises the foreign renewal price relative to a comparison with the same foreign marginal cost. Across dates or technology states, the foreign price rises exactly when
\begin{equation}
 \frac{\zeta_j}{1+t}\Delta\!\left(\frac{w}{h_j(H_j)}\right)
          +(1-\zeta_j)\Delta c_{Aj}>0.
 \label{eq:renewal_condition}
\end{equation}
On the floor branch, the price effect per unit of capability is
\begin{equation}
 P_{jH}=-\frac{\zeta_j\underline w}{1+t}
                  \frac{h'_j(H_j)}{h_j(H_j)^2}<0.
 \label{eq:PH}
\end{equation}
\end{proposition}

Hold-up here is collective: each atomistic buyer takes domestic capability as given, while aggregate sourcing decisions change the fallback available to all buyers at renewal. Holding installed procurement quantities fixed, a reduction in future capability creates a repricing exposure
\begin{equation}
 \mathcal W_H=\vartheta\sum_j q_{j1}P_{jH}\,\dd H_{j1}>0
 \quad\text{when }\dd H_{j1}<0,
 \label{eq:repricingwedge}
\end{equation}
where $\vartheta>0$ is a discount factor. A preservation increment has the opposite sign and avoids this price component. This foreign-price effect applies under the stated bargaining regime and at the wage floor. The complete comparison also includes domestic productivity, quantity adjustment, and the incidence of taxes and ownership.

Provider bargaining power and capability are separate objects. The stock $H$ changes the fallback cost while $\zeta$ remains the bargaining weight. With $\zeta=0$, capability loss has no effect on the foreign quoted price under this protocol. With no capability loss, a positive $\zeta$ distributes the current saving but does not by itself generate rising renewal rents.

The original cost advantage must continue to be checked. Capability loss can leave agent execution cheaper than rebuilding locally while making both options more expensive. The equilibrium formulas from Section 3 still apply at each date with the relevant $h$ and $P$, but the sign conditions must be evaluated again. Higher foreign prices can restore demand for human execution if a sourcing boundary or an alternative supplier becomes binding. A price increase is an adverse income terms-of-trade change; it does not automatically lower reference-price production volume.

The domestic fallback is the relevant bargaining alternative when it is cheaper than switching to another foreign supplier. If an effective rival price, including migration cost, is $\widehat P_j$, a simple competitive constraint is
\begin{equation}
 P_j^{\rm eff}=\min\{P_j^{\rm Nash},\widehat P_j\}.
 \label{eq:rival}
\end{equation}
When the rival cap binds, additional domestic capability loss has no local effect on the foreign price. Portability and viable alternative suppliers can therefore protect the importing economy even when domestic execution has contracted. Claims of hold-up require evidence about these alternatives as well as provider concentration.

A constant local-service import bill can coexist with less effective execution and a greater foreign rent share. Appendix \ref{app:renewal_details} derives that comparison, nets introductory concessions against later payments, and reports partial tax incidence.

\input{quantitative_results}

\clearpage
\section{Conclusions}
Using foreign agents in local services raises the tradable expansion needed to sustain domestic income. At the reservation wage, income falls when the local-service task frontier advances without an offset elsewhere. Output per paid hour still rises. Whether residents can consume more depends on how much of the technological saving reaches them through lower prices.

Supplier bargaining affects local services and tradables differently. Once wages hit the floor, relatively little supplier power can make local-service agentification lower value added at constant prices. In tradables it takes much more, because production can expand and earn export revenue. The finite compensation conditions distinguish the export growth needed to preserve income from that needed to preserve consumption.

The supply response can change the aggregate outcome. Replacing the Mexican nonlabor-share proxy with the more elastic external capacity benchmark expands the range of supplier terms under which consumption rises. A greater initial distance from the reservation wage allows more adjustment through wages before paid hours contract. Cheaper agents help consumers, while a lower export price raises the physical output needed to pay foreign suppliers.

There is also a slower risk. If human execution shrinks enough to weaken the option of returning to local production, foreign suppliers can raise renewal prices. Keeping an alternative viable can restrain those prices. For an agent-importing economy, evaluating adoption by a firm's cost saving or its output per paid hour alone misses the export earnings needed to finance that execution.

\clearpage
\appendix
\section*{Appendices}
\addcontentsline{toc}{section}{Appendices}
\section{Derivations and proofs}
\label{app:proofs}
\subsection{Task cost minimization}
For a fixed task aggregate $X_j=\exp\{\int\log x_j(i)\dd i\}$, unit cost is the geometric mean $G_j$ of task costs. At the optimum each task receives the same expenditure. Maximizing $p_jA_jK_j^{\alpha_j}X_j^{\beta_j}-G_jX_j$ gives $G_jX_j=\beta_j\R_j$. Hence human task quantities are $x_j^H=\beta_j\R_j/(w/h_j)$ and agent quantities $x_j^A=\beta_j\R_j/[(1+t)P_j]$, per unit measure. Integrating yields \eqref{eq:shares}. Substitution into production gives \eqref{eq:supply}. Euler's relation gives the fixed-factor payment $\alpha_j\R_j$. The residual sector is the same calculation with human execution only and $h_R=1$.

\subsection{Aggregation and fiscal neutrality}
Aggregate household expenditure at consumer prices is
\begin{align*}
 M&=(1-\tau_L)wL+\Pi+B_U+b(\bar L-L)+\mathcal S+F\\
  &=wL+\Pi+tI+\tau_VE+F=V+F+\tau_VE=(1+\tau_V)E.
\end{align*}
External financing is received and spent by residents in this accounting extension; the main model sets it to zero. Identical Cobb--Douglas weights imply $(1+\tau_V)p_jC_j=\gamma_jM$. Thus $p_jC_j=\gamma_jE$. The maintained nominal threshold compares $(1-\tau_L)w/[(1+\tau_V)\PC]$ with $r/[(1+\tau_V)\PC]$. Hence participation in \eqref{eq:floor} is unchanged. The nonfiscal reservation component does not enter $M$ as spending income. This step uses the nominal-threshold closure, separately from the demand aggregation. Aggregation therefore removes $\tau_V$ from goods and labor clearing. Individual budgets and the feasibility bound remain relevant.

\subsection{Proof of Proposition \ref{prop:allocation}}
From \eqref{eq:external}, $\gamma_TE+\mu_TY_T+\mu_N\gamma_NE=Y_T$ at $F=0$. Rearranging gives $E=kY_T$. Summing \eqref{eq:shares} and \eqref{eq:residual} then gives $wL=s_LE$. Local prices are uniquely recovered from $\R_N=\gamma_NE$, $\R_R=\gamma_RE$, and their production equations. Explicitly,
\begin{equation}
 p_N=\frac{(\gamma_NE)^{\alpha_N}G_N^{\beta_N}}
                 {A_NK_N^{\alpha_N}\beta_N^{\beta_N}},\qquad
 p_R=\frac{(\gamma_RE)^{\alpha_R}w^{\beta_R}}
                 {A_RK_R^{\alpha_R}\beta_R^{\beta_R}}.
 \label{eq:local_prices_closed}
\end{equation}

Under \eqref{eq:bargain}, $C_T^A=\zeta_Tw/h_T+(1-\zeta_T)(1+t)c_{AT}$ is increasing in $w$. Therefore $G_T$ increases with $w$ and $Y_T(w)$ decreases strictly. Since $s_Lk>0$, $L^d(w)=s_LkY_T(w)/w$ decreases strictly and tends to zero as $w\to\infty$. Complementarity determines the unique floor or full-employment branch. The strict eligibility condition is valid at $\underline w$ and becomes weaker at higher wages.

For a fixed posted foreign price rather than \eqref{eq:bargain}, a useful closed form is $Y_T=Z_Tw^{-\epsilon_T}$, where
\begin{equation}
 \epsilon_T=\frac{\beta_T(1-m_T)}{\alpha_T},\qquad
 Z_T=K_TA_T^{1/\alpha_T}
  \left(\frac{\beta_Th_T^{1-m_T}}{[(1+t)P_T]^{m_T}}\right)^{\beta_T/\alpha_T}.
\end{equation}
Then $w^{FE}=(s_LkZ_T/\bar L)^{1/(1+\epsilon_T)}$ and $w=\max\{\underline w,w^{FE}\}$. The bargained-price baseline uses the monotone scalar equation instead of this posted-price shortcut.

Differentiate \eqref{eq:sl}, using $\partial\mu_j/\partial m_j=\beta_j/(1+t)$ and $\partial\ell_j/\partial m_j=-\beta_j$. This gives \eqref{eq:sl_derivatives}. From $w\bar L=s_LE$ on the full-employment branch, $E/w=\bar L/s_L$ and \eqref{eq:residual_absorb} follows. The same wage-bill identity gives \eqref{eq:wagefall}. At the floor, $L=s_LE/\underline w$ and $L_R=\beta_R\gamma_RE/\underline w$. If $E$ falls and $s_L$ weakly falls, both total and residual paid hours fall.

\input{joint_proofs}

\subsection{Proof of Proposition \ref{prop:threshold} and the volume identity}
At unchanged floor wage, fixed $K,A,h$, the supply equation for $T$ gives $r_T=g_T^{-\beta_T/\alpha_T}$. Using the multiplier in both equilibria gives \eqref{eq:finiteE}. For $N$, production at given gross revenue implies $Y_N=A_NK_N^{\alpha_N}(\beta_N\gamma_NE/G_N)^{\beta_N}$, giving \eqref{eq:finiteY}. The residual equation gives $r_R=\mathcal E^{\beta_R}$. Human labor ratios are $L_{T1}/L_{T0}=r_T(1-m_{T1})/(1-m_{T0})$ and $L_{N1}/L_{N0}=\mathcal E(1-m_{N1})/(1-m_{N0})$. Dividing output ratios by these labor ratios proves the productivity expressions. Proposition \ref{prop:allocation} supplies the aggregate-hours sign. Balanced trade gives $C_{T1}/C_{T0}=\mathcal E$. Combining this with the two local output ratios yields $Q_1/Q_0=\mathcal E^B g_N^{-\beta_N\gamma_N}$. Substituting $\mathcal E=r_T/\mathcal H_E$ proves \eqref{eq:consumption_hurdle} and its threshold equivalence.

For the local volume identity, the production envelope at fixed domestic complements and human efficiency is
\begin{equation}
 \sum_jp_j\dd Y_j
   =w\,\dd L+(1+t)\sum_{j=T,N}P_j\dd q_j
       +\sum_{j=T,N}\beta_j\R_j\delta_j\dd m_j.
 \label{eq:primal_envelope}
\end{equation}
One can also verify this directly using $\dd\log G_j=-\delta_j\dd m_j+m_j\dd\log P_j$ at the floor and $\dd\log q_j=\dd\log\R_j+\dd m_j/m_j-\dd\log P_j$. Subtracting $\sum P_j\dd q_j$ yields \eqref{eq:volume_envelope}. The tax term uses quantity changes valued at the current foreign price; it is not $t\dd I$ when $P$ changes. This distinction matters for the numerical check.

\subsection{External resources and consumption}
Equation \eqref{eq:netexports} follows from external clearing. With a fixed nonzero $F$, rearranging the same equation yields \eqref{eq:withF}. Restoring $E_0$ at the new technology on the floor branch leaves $Y_{T1}$ unchanged and gives $F^*=\Xi_1E_0-(1-\mu_{T1})Y_{T1}=\Xi_1(E_0-E_1)$. It is valid only if the restored-demand allocation stays on that branch and remains fiscally feasible.

At $F=0$, differentiate $V=E=\sum_jp_jY_j-\sum_jP_jq_j$ to obtain
\begin{equation}
 \dd E=\dd V^Q+Y_N\dd p_N+Y_R\dd p_R-\sum_{j=T,N}q_j\dd P_j.
\end{equation}
Since $\dd\log\PC=\gamma_N\dd\log p_N+\gamma_R\dd\log p_R$ and $p_jY_j=\gamma_jE$ in the local sectors, subtracting $E\dd\log\PC$ proves \eqref{eq:realconsumption}. The quantity index $Q$ is an aggregate consumption object. A welfare ranking with different household weights or nonmarket costs requires additional structure.

\subsection{Unit bargaining and proof of Proposition \ref{prop:renewal}}
\label{app:bargain}
For one required execution unit, supplier surplus is $P-c_A$ and the domestic procurement saving is $C^H-(1+t)P$. The unit protocol chooses $P$ to maximize
\begin{equation}
 (P-c_A)^\zeta[C^H-(1+t)P]^{1-\zeta}
\end{equation}
on the nonempty interval $c_A<P<C^H/(1+t)$. Its first-order condition is
\begin{equation}
 \frac{\zeta}{P-c_A}
  =\frac{(1-\zeta)(1+t)}{C^H-(1+t)P},
\end{equation}
which gives \eqref{eq:bargain}; the log objective is strictly concave. Substitution of $C^H=w/h(H)$ and finite differencing gives \eqref{eq:renewal_condition}. The outside-option condition follows from the log ratio. Differentiation at the floor yields \eqref{eq:PH}. Comparing the prices at $t$ and zero, holding the outside option fixed, proves \eqref{eq:taxincidence}. These calculations describe the unit protocol; arbitrary supplier demand-based pricing would require another price-setting model.

For the comparison in \eqref{eq:renewalQ}, the task shares and tradable cost are unchanged, so the multiplier leaves $E$ fixed. Labor demand depends on revenues and the fixed task shares, hence paid hours are unchanged. The foreign nontradable bill is $\mu_N\gamma_NE$, so $\dd\log q_N=-\dd\log P_N$. Differentiating the price rule gives $\dd\log P_N=-\theta_N\dd\log h_N$. At fixed $E$, production gives $\dd\log Y_N=\ell_N\dd\log h_N-\beta_Nm_N\dd\log P_N$. Only nontradable task output changes in the final consumption bundle; multiplying by $\gamma_N$ proves \eqref{eq:renewalQ}. Finally, with $I_N$ and $c_{AN}$ fixed, $\dd\Pi_N^F=-c_{AN}\dd q_N=c_{AN}q_N\dd\log P_N>0$.

\input{export_price_proofs}

\input{fiscal_details}

\input{renewal_details}

\input{finite_bounds}

\input{ces_robustness}

\input{quantitative_appendix}
\FloatBarrier
\input{observable_markers}

\section{Public external liquidity}
\label{app:liquidity}
The balanced-trade model describes the real adjustment when external financing does not expand. To discuss reserves, let an exogenous private inflow be $\bar F_s$ and permit a public draw $R_s-R_{s+1}$, so that $F_s=\bar F_s+R_s-R_{s+1}$. A draw fills a current resource gap and lowers the next reserve stock. If the social continuation value is $\Psi_s(R_{s+1})$ with $\Psi_s'>0$, the marginal external-liquidity cost of that draw is $\lambda_s^{FX}=\Psi_s'(R_{s+1})$.

For a given production allocation, an additional net import requirement $\dd I_s-\dd Y_{Ts}+\dd C_{Ts}$ has a liquidity cost only to the extent that reserves absorb it. This reduced-form public cost is separate from the foreign supplier rent and from the domestic production volume comparison. A private adopter need not internalize the marginal value of the common reserve buffer \citep{Lorenzoni2008,JeanneKorinek2010,Bianchi2011,JeanneRanciere2011}.

An early concession and a later payment can have the same discounted private value while differing in their liquidity value if $\lambda_0^{FX}$ and $\lambda_1^{FX}$ differ. In that case lifecycle rent netting does not guarantee liquidity neutrality. This extension supplies a two-date accounting and shadow-value interpretation. It does not solve reserve accumulation, crisis probabilities, sovereign borrowing, or an intertemporal optimum. Those objects require a further closure.

\section{Scope conditions and limiting cases}
\label{app:scope}
The main production comparisons condition on a nominal reservation-wage floor and initially fixed human execution efficiency. They retain an endogenous tradable supply response and local demand adjustment. They assume perfect substitution within eligible tasks, Cobb--Douglas aggregation across tasks and consumption goods, positive resident-owned fixed factors, no new human tasks, and a fixed external-financing closure. These restrictions make the aggregate account tractable. They do not establish an empirical prevalence of substitution or an inevitable decline under agentification.

The results extend locally to small perturbations that preserve strict signs and a regular equilibrium. This continuity statement does not provide a quantitative robustness region. Large departures from the assumed task structure, household expenditure shares, or ownership require solving the modified model. A fixed real consumption floor, rather than a numeraire floor, would remove the fiscal neutrality used here. Foreign ownership of the fixed factors would introduce additional external income payments.

With $m_T,m_N\to1$ at positive execution prices, labor remains in $R$ when $\gamma_R>0$. Domestic fixed-factor income also remains. The economy need not lose all domestic value added. If the residual share is also taken to zero, paid labor can vanish and the public support bill approaches $B_U+b\bar L$. Its feasibility then depends on the remaining domestic rent income and agent-tax receipts. At $t=0$ the required consumption tax in that limit is $(B_U+b\bar L)/E$. Whether it exceeds an admissible maximum depends on $E$; it is not an accounting inevitability.

Very cheap imported execution can expand tradable supply enough to reverse the adverse result. A low tradable fixed-factor exponent makes that response especially elastic. The inverse case, foreign agents priced close to effective domestic execution cost, transfers a larger task expenditure share abroad while providing little operating cost relief. These limits explain why supplier terms and tradable scalability matter jointly.

The two-date capability extension specifies how execution sustains a shared state. It does not establish an infinitely persistent loss, multiple steady states, or a global transition. A reduction in sectoral human hours lowers capability relative to a comparison path; it need not reduce it below its initial level if the remaining execution offsets depreciation. The bargaining effect can be capped by rival supply, and falling foreign marginal cost can dominate renewal pressure. Restoring $h$ or adding human tasks can change the sourcing regime.

\section{Notation and reproducibility}
\label{app:notation}
\begin{longtable}{p{.19\textwidth}p{.74\textwidth}}
\toprule
Symbol & Meaning\\\midrule\endhead
$T,N,R$ & Tradable task sector, nontradable task sector, human-only residual activity.\\
$\alpha_j,\beta_j$ & Fixed-factor and execution shares; $\beta_j=1-\alpha_j$.\\
$\gamma_j$ & Common final-expenditure weight.\\
$\eta_j,m_j$ & Eligible task frontier and actual imported execution share of tasks.\\
$w,\underline w,L,\bar L$ & Gross wage, reservation wage, paid hours, available labor capacity.\\
$r,d_0$ & Net reservation threshold and initial proportional wage distance to it.\\
$B_U,b$ & Aggregate universal support and payment per unused capacity-hour.\\
$\kappa,\rho,\Gamma,J,\mathcal J_j$ & Public benefit/wage; fiscal wage coefficient; direct fiscal coefficient; import/expenditure ratio; its frontier derivatives.\\
$\tau_L,t,\tau_V$ & Labor, agent-purchase, and final-consumption taxes.\\
$P_j,c_{Aj},\zeta_j$ & Foreign unit price, marginal cost, and provider bargaining weight.\\
$h_j,H_j$ & Human execution productivity and the shared capability that supports it.\\
$G_j,\delta_j$ & Geometric execution cost and log saving from eligible agent execution.\\
$\R_j,\ell_j,\mu_j$ & Gross sector revenue, human wage share of revenue, foreign import share of revenue.\\
$E,V,Y^D,V^Q$ & Pretax final expenditure, national resources excluding final VAT, domestic factor income, reference-price value added.\\
$\mathcal E$ & Expenditure ratio $E_1/E_0$ in finite comparisons.\\
$B,\mathcal A_j$ & Composite expenditure weight and sectoral comparative-static coefficients.\\
$\mathcal H_E,\mathcal H_Q$ & Gross tradable output ratios required to preserve income and consumption on the floor branch.\\
$p_T,\varpi,p_T^*$ & Export price in external units; its finite ratio; price at which the reservation wage binds.\\
$\varepsilon_T^S,\omega_j,\psi_T$ & Conditional tradable-supply elasticity; execution-cost wage elasticity; full-employment wage elasticity to the export price.\\
$\overline\delta_j,\phi_j$ & Frontier sign bounds and foreign-cost pass-through at the floor.\\
$\Xi,s_L,s_K$ & Import-leakage denominator, wage-bill share of $E$, fixed-factor rent share of $E$.\\
$I,X,F$ & Agent imports, net tradable exports, net external financing.\\
$\PC,Q$ & Producer-price final-consumption index and aggregate real consumption.\\
$\mathcal S,\bar\tau_V$ & Surplus rebate to owners and maximum admissible VAT.\\
$R_s,\lambda_s^{FX}$ & Public reserves and their marginal social continuation value.\\
\bottomrule
\end{longtable}

The source archive contains the equilibrium solver \path{model.py}, the trajectory runner \path{simulate_paths.py}, the grid mapper \path{grid_maps.py}, the accounting bridge \path{anchors.py}, and nine verification scripts with machine-readable reports. The checks reconstruct task production from human and agent execution quantities, compare the reduced allocation with primitive market equations in both labor regimes, verify household and government budgets, and compare finite changes and numerical derivatives with the stated formulas. Symbolic checks cover the sectoral coefficients, finite bargaining bounds, unit bargaining, and partial tax incidence. Synthetic inputs verify implementation. A separate CES check reconstructs task production and independent market clearing, recovers the benchmark limit, and verifies the effective-gap decomposition and strict adverse regions. Separate reservation/fiscal checks verify independent public support, the fiscal-tax bound and its reversal, the nominal-threshold neutrality restriction, and the wage-distance diagnostics. A transfer-rule verifier checks universal and contingent support, exact finite fiscal changes, and budgets in both labor regimes, including CES tasks and export-price changes. Grid checks additionally cover paid-price admissibility, a common initial allocation across bargaining weights, fiscal failure and subsequent recovery, and plot export. A seventh verifier checks the extracted country source cells, moment reconstruction, initial reference prices, and the finite consumption export requirement against solved production primitives. An eighth verifier tests the external-numeraire export-price extension against independent primitive equilibria, both regime derivatives, finite compensation, and the exact reservation-wage crossing. The one-command reproduction entry point is \path{reproduce.py}; \path{REPLICATION.md} maps every table and figure to its producer and inputs. The country exercise reports structural sensitivity with historical accounts and unestimated technology and supplier axes. Post-boundary diagnostic allocations are labeled separately from financed paths.

\bibliographystyle{apalike}
\begingroup
\small
\setstretch{1.0}
\setlength{\bibsep}{5pt}
\bibliography{references}
\endgroup
\end{document}

%% file: introduction.tex
\section{Introduction}
An accounting firm can serve more clients per paid hour by delegating execution to a foreign AI agent. Its customers remain domestic, but part of what they pay now remunerates a supplier abroad. Customer support, software, and back-office work can follow the same path: a business that sells at home starts importing part of what it produces.

We study an economy with a tradable sector, local services that can use agents, and a human-only residual activity. Firms adopt agents when they cost less than domestic execution. Workers move between activities, and wages adjust until they reach a reservation threshold. Further displacement can then reduce paid hours. Supplier bargaining determines how much of the technological saving reaches the buyer.

At the reservation wage, delegating more local-service tasks to foreign agents lowers current income. Output per paid hour rises, yet value added at constant prices can fall. Its sign depends on how much of the saving the buyer keeps. At the Brazilian reference allocation, a supplier bargaining share above about 0.15 is enough for a marginal increase in local services' use of foreign agents to lower value added at constant prices, however low suppliers' costs are. In tradables, the corresponding threshold is about 0.66. The gap comes from the export response: tradable firms can expand production and earn foreign exchange to help pay for agents.

Foreign agents must be paid for. Let $E$ denote pretax final expenditure, $Y_T$ gross tradable output, $\gamma_j$ expenditure weights, and $\mu_j$ foreign-agent payments as a share of sectoral revenue. With the tradable price normalized to one and balanced trade,
\begin{equation}
 E=\frac{(1-\mu_T)Y_T}{\gamma_T+\gamma_N\mu_N}.
 \label{eq:intro_multiplier}
\end{equation}
Greater use of foreign agents in local services raises the denominator. Using agents in tradable production reduces the share of tradable revenue retained domestically, but can also raise output. We derive how much tradable production must grow to preserve income and whether lower execution costs generate that growth.

Consumption can rise even when income falls. Cheaper agents lower the price of local services, allowing residents to buy more with less domestic income. We therefore derive separate export requirements for preserving income and real consumption. Supplier bargaining limits this price gain. A lower export price raises both requirements and can push wages down to the reservation threshold, after which paid hours contract.

The quantitative exercise uses pre-generative-AI expenditure and factor-income accounts for Brazil and Mexico. Replacing the nonlabor-share proxy with a more elastic tradable supply response enlarges the consumption-gain region, especially for Mexico. Allowing wages to fall further before reaching the reservation threshold also shifts the boundary. CES task aggregation preserves both gains and losses. The accounts set the reference economic structure. We do not observe task replacement, supplier costs, or bargaining shares, so the maps vary them.

Two extensions address public support and renewal contracts. A uniform consumption tax finances universal grants and a supplement for unused hours: income loss raises the burden of the universal component, while displacement also expands the contingent bill. Domestic execution also sustains capabilities that buyers may need if they return to human production. When those capabilities erode, the cost of returning rises and foreign suppliers can demand higher renewal prices.

The closest paper is \citet{AtheyScottMorton2025}. They study how pricing by foreign AI suppliers shapes wages and welfare, including the effects of usage and access fees on entry and variety. We compare agentification in tradables and local services within a common equilibrium. At the reservation wage, local-service agentification requires a larger retained cost saving than tradable agentification to raise domestic value added. Agent imports in local services must be financed by tradable production; agentification in tradables can itself expand that financing base. We derive separate export-growth requirements for preserving income and consumption. We also formalize the capability-loss concern in their research agenda: aggregate sourcing decisions change the common human fallback and, through it, renewal prices. Our unit-pricing limit connects to their zero-access-fee adoption boundary.

\citet{Brecher1974} provides the trade benchmark with a wage floor and unused labor. \citet{GilbertOladi2021} show that labor-eliminating technology can reduce output and welfare with a manufacturing wage distortion. We trace the imported-input channel when substitution reaches local services: foreign agents must be paid through the external account, and supplier bargaining determines the saving retained domestically. Our net reservation threshold $r$ applies across activities, and we ask what happens once it binds. Its nonfiscal component $r-b$ is an earnings-equivalent reservation term, absent from measured value added and market consumption. \citet{AcemogluRestrepo2026} study inefficient automation through the dissipation of labor rents above workers' outside options.

\citet{MontesRojasEtAl2026} study imported AI in a formal tradable sector, with workers moving to informal nontradable production and external debt responding through a country-risk premium. Here agents also execute tasks in local services, and consumption adjusts to a fixed external-financing closure. \citet{KorinekStiglitz2021} and \citet{AlonsoEtAl2022} examine how automation can worsen developing countries' international position. We use the task structure of \citet{AcemogluRestrepo2019} and retain the gains from cheaper imported inputs emphasized by \citet{GrossmanRossiHansberg2008,GoldbergEtAl2010,HalpernKorenSzeidl2015}.

Sections 2 and 3 develop the model and production results. Section 4 studies renewal contracts, Section 5 reports the quantitative exercise, and Section 6 concludes.

%% file: joint_results.tex
\subsection{Sectoral agentification, value added, and real consumption}
\label{sec:volume}
To distinguish production quantities from income, define value added at initial reference prices:
\begin{equation}
 V^Q_s=\sum_jp_{j0}Y_{js}-\sum_{j=T,N}P_{j0}q_{js},\qquad s=0,1.
 \label{eq:VQ}
\end{equation}
At the reference equilibrium, $V^Q_0=V_0=E_0$. At $t=0$, this is domestic factor value added; with positive fixed agent taxes it includes their consistent reference-price valuation. The final-consumption price and quantity indices are
\begin{equation}
 \PC=\prod_j(p_j/\gamma_j)^{\gamma_j},\qquad
 Q=\frac{E}{\PC}=\prod_jC_j^{\gamma_j}.
 \label{eq:Q}
\end{equation}
Balanced trade links real consumption directly to production and foreign input prices:
\begin{equation}
 E\,\dd\log Q=\dd V^Q-\sum_{j=T,N}q_j\,\dd P_j.
 \label{eq:realconsumption}
\end{equation}
At unchanged foreign prices, a local fall in value added at constant prices reduces real consumption by the same amount in expenditure units. Falling agent prices give consumers a terms-of-trade gain.

For the local comparison, let the floor bind with strictly positive nonemployment, hold domestic complements and human efficiency fixed, and initially set $t=0$. Define
\begin{align}
 \delta_j&=\log\left(\frac{\underline w/h_j}{P_j}\right)>0,
 \label{eq:delta}\\
 B&=\gamma_T+\beta_N\gamma_N+\beta_R\gamma_R,
 \qquad
 \mathcal A_T=\frac{\beta_TB}{\alpha_T},\quad
 \mathcal A_N=\beta_N\gamma_N,
 \label{eq:joint_coeffs}\\
 \overline\delta_T&=\frac{\alpha_T}{1-\mu_T},\qquad
 \overline\delta_N=\frac{B}{\Xi}
      =\frac{s_L}{\Xi}+\frac{\alpha_T}{1-\mu_T}.
 \label{eq:joint_bounds}
\end{align}
The coefficients $\mathcal A_j$ are comparative-static weights, distinct from the productivity parameters $A_j$. The bounds depend on equilibrium task shares. They are evaluated at the reference allocation.

\begin{proposition}[Sectoral frontiers and supplier bargaining]
\label{prop:joint}
At a fiscally feasible interior floor-branch equilibrium with $t=0$, hold foreign marginal costs, human capability, and bargaining weights fixed. For any simultaneous frontier changes $\dd m_T,\dd m_N\ge0$,
\begin{align}
 \frac{\dd V^Q}{E}
 &=\mathcal A_T(\delta_T-\overline\delta_T)\dd m_T
   +\mathcal A_N(\delta_N-\overline\delta_N)\dd m_N,
 \label{eq:jointV}\\
 \dd\log E
 &=\frac{\beta_T}{\alpha_T}(\delta_T-\overline\delta_T)\dd m_T
      -\frac{\beta_N\gamma_N}{\Xi}\dd m_N.
 \label{eq:jointE}
\end{align}
Nontradable frontier expansion always lowers current income in this comparison. It lowers volume value added when $\delta_N<\overline\delta_N$. Tradable frontier expansion lowers both when $\delta_T<\overline\delta_T$. The bargaining rule implies $0<\delta_j<-\log\zeta_j$, so
\begin{equation}
 \zeta_j>\exp(-\overline\delta_j)
 \label{eq:joint_zeta}
\end{equation}
is sufficient to make sector $j$'s frontier contribution adverse at every eligible foreign marginal cost level on the maintained branch. If this condition holds in both sectors, every nonzero nonnegative mixture of frontier expansion locally lowers income, reference-price value added, real consumption, and paid hours. The two bargaining thresholds satisfy
\begin{equation}
 \overline\delta_N>1>\overline\delta_T,
 \qquad
 \exp(-\overline\delta_N)<\exp(-1)<\exp(-\overline\delta_T).
 \label{eq:joint_ordering}
\end{equation}
\end{proposition}

The first level of the result concerns local services. Their frontier expansion raises the imported content of local demand while leaving tradable supply unchanged at the floor. It always lowers current income in this comparison. A sufficient provider share makes its volume contribution negative even when foreign execution is technologically very cheap. The second level concerns tradables. Their expansion generates gross output and foreign exchange; ruling out a favorable contribution even at very low supplier costs requires a higher bargaining weight. A weight below that sufficient threshold leaves the sign dependent on the actual cost gap.

The asymmetry orders the critical bargaining weights. The parameter $\zeta_j$ is the supplier's share of unit procurement surplus under the maintained protocol. Its numerical size alone identifies neither a market share nor a price--cost markup. The same supplier weight can eliminate the favorable nontradable frontier effect while still allowing a favorable tradable response.

The gap between private cost savings and aggregate gains has a policy implication for an agent-importing economy. A positive saving makes foreign execution attractive to an individual firm. At the reservation wage and unchanged foreign prices, the saving required for a positive aggregate volume and consumption contribution is larger in local services than in tradables. This distinction supplies a criterion for evaluating adoption incentives and public procurement: the private saving must be considered together with domestic value retention, foreign-exchange generation, and the price paid to the supplier. Policies that improve effective supplier terms or expand tradable capacity strengthen the consumption comparison under this closure. The bounds $\overline\delta_j$ are aggregate sign conditions. Deriving a planner's adoption rule or a corrective tax additionally requires nonmarket valuations and policy incidence. The mechanical labor--agent tax alignment in Section \ref{sec:fiscal} addresses a separate margin.

The tradable offset is explicit. If $\delta_T>\overline\delta_T$, $\delta_N<\overline\delta_N$, and $\dd m_N>0$, volume value added falls precisely when
\begin{equation}
 \frac{\dd m_T}{\dd m_N}
 <\frac{\mathcal A_N(\overline\delta_N-\delta_N)}
         {\mathcal A_T(\delta_T-\overline\delta_T)}.
 \label{eq:frontier_mix}
\end{equation}
Both frontiers can advance by economically relevant relative amounts. The formula includes the endogenous expansion of tradable output at the stated relative frontier speeds.

Higher paid-hour productivity can also coexist with higher gross output in both task sectors. For nontradable expansion alone, let $c_N=\beta_N\gamma_N/\Xi$. Gross nontradable output rises if $\delta_N>c_N$, and
\begin{equation}
 \overline\delta_N-c_N
   =\frac{\gamma_T+\beta_R\gamma_R}{\Xi}>0.
 \label{eq:intervalpositive}
\end{equation}
The overlap $c_N<\delta_N<\overline\delta_N$ exists under bargaining when $\zeta_N<\exp(-c_N)$. With joint frontier changes, gross nontradable output rises when $\dd\log E+\delta_N\dd m_N>0$, while gross tradable output rises whenever $\dd m_T>0$. These are additional output conditions; paid-hour productivity requires less.

\subsection{Frontier expansion and falling foreign costs}
\label{sec:pricespeeds}
The third level of the result concerns net real consumption when agents become cheaper as their task range expands. With the floor and human efficiency fixed, define
\begin{equation}
 \mathcal C_T=\frac{s_L\beta_Tm_T}{\alpha_T},\qquad
 \phi_j=\frac{(1-\zeta_j)c_{Aj}}{P_j},\qquad
 \dd\log P_j=\phi_j\dd\log c_{Aj}.
 \label{eq:passthrough}
\end{equation}
At $t=0$, the exact local production and consumption comparisons are
\begin{align}
 \frac{\dd V^Q}{E}
 &=\sum_{j=T,N}\mathcal A_j(\delta_j-\overline\delta_j)\dd m_j
                               -\mathcal C_T\dd\log P_T,
 \label{eq:joint_costV}\\
 \dd\log Q
 &=\sum_{j=T,N}\mathcal A_j
       \bigl[(\delta_j-\overline\delta_j)\dd m_j
                         -m_j\dd\log P_j\bigr].
 \label{eq:joint_costQ}
\end{align}
When both frontier terms are adverse, let
\begin{equation}
 \mathcal D=\sum_j\mathcal A_j(\overline\delta_j-\delta_j)\dd m_j>0.
\end{equation}
For reductions in foreign prices, volume value added falls if and only if $\mathcal D>\mathcal C_T(-\dd\log P_T)$. Real consumption falls if and only if
\begin{equation}
 \mathcal D>
 \mathcal A_Tm_T(-\dd\log P_T)
   +\mathcal A_Nm_N(-\dd\log P_N).
 \label{eq:consumption_speed}
\end{equation}
A nontradable price reduction alone leaves expenditure, paid hours, and local reference-price value added unchanged on this branch while increasing real consumption. A tradable price reduction can also expand production and paid hours. The net consumption effect depends on both task expansion and the price decline.

Bargaining limits the pass-through: $0<\phi_j<1-\zeta_j$, and $\phi_j$ tends to zero as foreign marginal cost tends to zero. Under the strict bargaining thresholds, a sufficient uniform condition for consumption contraction is
\begin{equation}
 \sum_j\mathcal A_j(\overline\delta_j+\log\zeta_j)\dd m_j
 >\sum_j\mathcal A_jm_j(1-\zeta_j)(-\dd\log c_{Aj}).
 \label{eq:uniform_speed}
\end{equation}
Consumption depends on how quickly agent use expands relative to the decline in the price buyers pay.

The reservation-wage floor supplies the nonemployment margin in these comparisons. With full employment, unchanged human efficiency, and $t=0$, the production envelope gives a positive local volume effect from cost-saving frontier expansion. In the adverse frontier region, removing the floor shifts the local adjustment toward a lower clearing wage. The benchmark takes a fixed nominal reservation threshold and accounts separately for the financing of public support. Its dependence on that labor-supply closure is part of the result.

%% file: export_prices.tex
\subsection{Export prices in an external numeraire}
\label{sec:export_prices}
An importer can face cheaper agents and a lower price for its own exports. To separate these effects, let $p_T>0$ vary in a common external numeraire, with $p_{T0}=1$. The monetary reservation threshold $r$, support parameters $B_U,b$, and foreign marginal costs are fixed in that unit for a pure export-price change. The wage floor remains $\underline w=r/(1-\tau_L)$. A threshold indexed to the consumption basket would be a different closure.

Balanced trade now implies
\begin{equation}
 E=\frac{p_T(1-\mu_T)Y_T}{\Xi},\qquad
 \mathcal H_E^{p}=\frac{p_{T0}}{p_{T1}}
     \frac{1-\mu_{T0}}{1-\mu_{T1}}\frac{\Xi_1}{\Xi_0}.
 \label{eq:price_hurdle}
\end{equation}
$E_1/E_0=r_T/\mathcal H_E^{p}$: a lower export price raises the physical expansion required to preserve income. When the same floor binds at both endpoints, writing $\varpi=p_{T1}/p_{T0}$ gives
\begin{equation}
 \frac{Q_1}{Q_0}=\left(\frac{r_T}{\mathcal H_E^{p}}\right)^B
       \varpi^{-\gamma_T}g_N^{-\beta_N\gamma_N},\qquad
 \mathcal H_Q^{p}=\mathcal H_E^{p}
       \varpi^{\gamma_T/B}g_N^{\beta_N\gamma_N/B}.
 \label{eq:price_consumption_hurdle}
\end{equation}
Since $B>\gamma_T$, an export-price fall also raises the consumption-preserving hurdle, holding the execution-cost change fixed.

The productive response is endogenous. Define the conditional tradable-supply elasticity $\varepsilon_T^S=\beta_T/\alpha_T$ and the wage elasticity of its execution cost $\omega_T=\partial\log G_T/\partial\log w$. With technology and foreign marginal costs fixed, the full-employment wage satisfies
\begin{equation}
 \frac{\partial\log w^{FE}}{\partial\log p_T}
   =\frac{1+\varepsilon_T^S}{1+\varepsilon_T^S\omega_T}>0.
 \label{eq:price_wage}
\end{equation}
A pure fall in $p_T$ therefore moves that wage toward the fixed nominal floor. Once the floor binds, $\partial\log E/\partial\log p_T=\partial\log L/\partial\log p_T=1/\alpha_T$. For $t=0$, the local volume response is $E^{-1}\partial V^Q/\partial\log p_T=s_L/\alpha_T$, and the consumption elasticity is $B/\alpha_T-\gamma_T>0$. Appendix \ref{app:export_prices} derives both labor regimes, the exact price at which the floor binds, and the net-export terms-of-trade identity. We take the export price as given and ask what happens if it falls.

%% file: quantitative_results.tex
\section{Export compensation and structural sensitivity}
\label{sec:empirical}
Agent adoption can expand exports while raising the imported content of local production. The maps compare that expansion with the foreign payments it must cover.

\subsection{Structural moments}
The maps are structural comparisons. Historical accounts set the reference expenditure and factor shares. Task coverage, foreign costs, and supplier shares are scenario axes. The maps show where gains and losses occur, not where Brazil or Mexico stand today.

We use the Brazilian and Mexican structures in the WIOD 2016 release, with 2014 as the main pre-generative-AI reference year and 2010 as a source-year check \citep{TimmerEtAl2015,WIOD2016}. The national input--output tables and the Socio-Economic Accounts share an industry classification. The latter adjust labor compensation for self-employment using documented country-specific procedures \citep{GoumaEtAl2018}.

Household final expenditure at basic prices, including direct imports, determines the three expenditure weights. Goods-producing industries enter $T$. Service industries with exports of at least ten percent of gross output also enter $T$, except for the activities assigned to $R$. Construction, utilities, real estate, public services, education, health, personal services, and household employment enter $R$; the remaining services enter $N$. The administrative-support block is conservatively assigned to $R$ in both countries. In Mexico this published block has large household expenditure while the real-estate row is zero. Its internal composition cannot be recovered from the tables used here. An alternative assignment puts the entire administrative block in $N$. Service-export cutoffs of five and twenty percent supply additional classification checks. Every industry assignment and source cell is retained in the replication files.

\begin{table}[htbp]
\centering
\caption{Structural moments used in the main comparison}
\label{tab:country_moments}
\input{tables/country_moments}
\begin{minipage}{.96\textwidth}\small
Source: WIOD 2016 release, reference year 2014. Expenditure shares use household purchases by industry at basic prices. The $\alpha_j$ are empirical proxies based on published capital compensation divided by capital plus adjusted labor compensation. Rounded expenditure shares can sum to 1.001. The nonlabor residual includes components beyond the competitive resident fixed-factor rent represented in the model.
\end{minipage}
\end{table}

Conventional software and other intermediate inputs are consolidated out. Agent use starts at zero. We target household expenditure and factor-income shares, leaving government consumption, investment, and the observed trade balance outside the calibration. Published nonlabor compensation proxies for the fixed-factor share $\alpha_j$. We vary tradable supply responsiveness separately below.

\subsection{The export requirement and the endogenous response}
\begin{figure}[htbp]
\centering
\includegraphics[width=\textwidth]{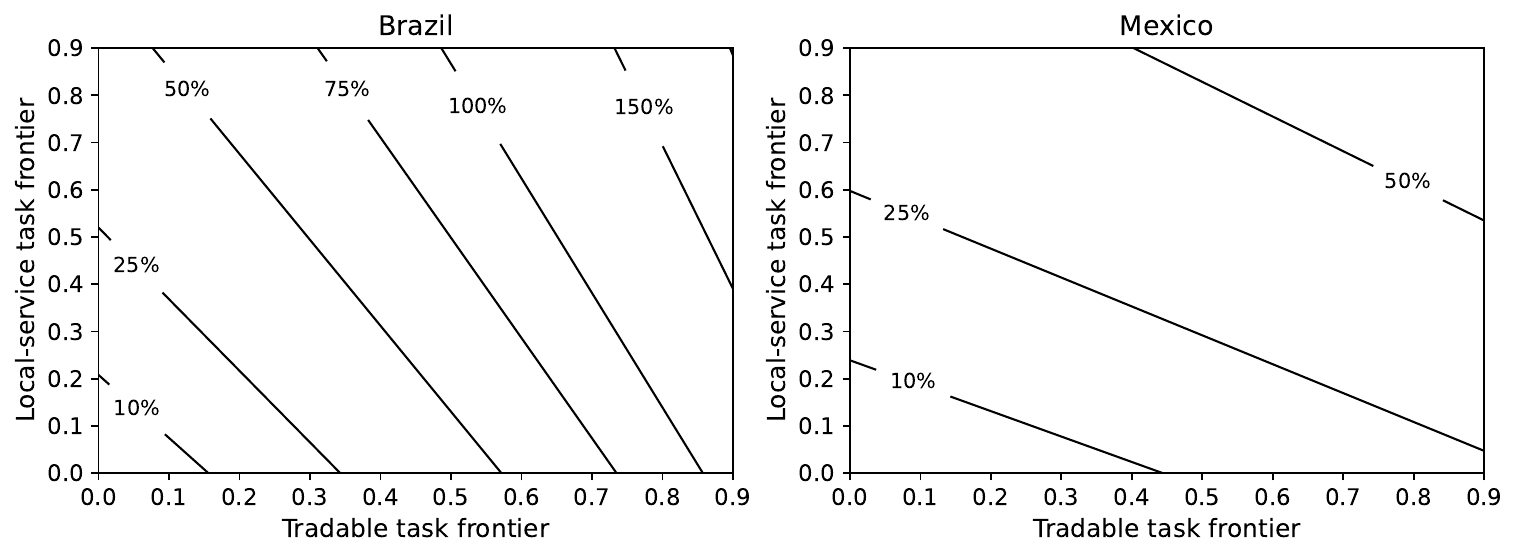}
\caption{Gross tradable expansion required to preserve national income. Contours report percent growth relative to the initial allocation. Both task frontiers vary; foreign costs and supplier power determine whether production can deliver the required expansion.}
\label{fig:export_hurdles}
\end{figure}

Figure \ref{fig:export_hurdles} reports the gross tradable growth required to preserve income. Starting from zero agent imports, this requirement is
\begin{equation}
 \mathcal H_E=\frac{1+(\gamma_N/\gamma_T)\beta_Nm_N}{1-\beta_Tm_T}.
 \label{eq:country_hurdle}
\end{equation}
The calculation spans both task frontiers. Greater local-service agentification raises the imported content of domestic demand; greater tradable agentification reduces the fraction of gross tradable revenue retained domestically. Tradable production can also grow as its execution cost falls. The relevant comparison is its endogenous expansion $r_T$ against this requirement. With the floor binding at both endpoints, equation \eqref{eq:consumption_hurdle} separately evaluates the smaller expansion needed to preserve real consumption when local execution becomes cheaper.

To compute the response, the initial allocation normalizes the wage and paid hours to one and chooses the supply scales to match that allocation. Modern agent execution is initially unavailable. Its foreign cost enters at the human-cost parity margin and subsequently falls by $g$ log points, so that $c_{Aj1}/w_0=\exp(-g)$, with $h_j=1$. A numerical offset of $10^{-7}$ makes agents strictly cheaper than human execution. The main maps vary the common supplier bargaining weight over $[0,.98]$ and $g$ over $[0,6]$. Separate panels extend both task frontiers to one quarter, one half, and three quarters of tasks. These frontier positions are substitution scenarios. The maximum cost decline explores the limiting case of very cheap execution. The comparison horizon is normalized, and $g$ is the total log cost decline.

\begin{figure}[htbp]
\centering
\includegraphics[width=\textwidth]{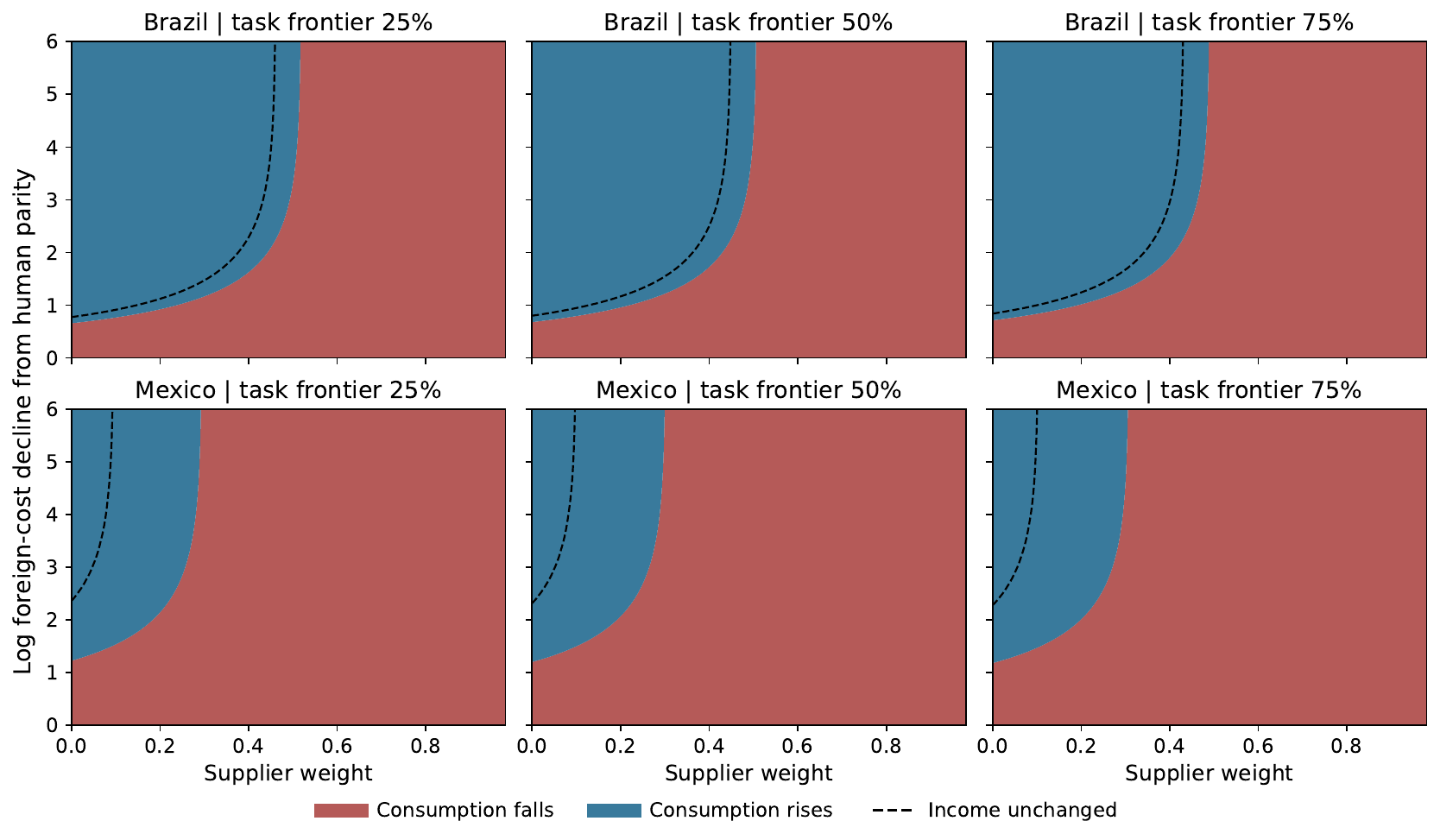}
\caption{Joint frontier expansion and foreign-cost decline. Color indicates the finite change in real consumption; dashed curves mark unchanged current income. Appendix \ref{app:quantitative} reports the initial-price value-added comparison across the admissible input reference-price range. Each point solves the labor regime endogenously.}
\label{fig:country_maps}
\end{figure}

Figure \ref{fig:country_maps} solves wages, production, import payments, local prices, and paid hours at each endpoint. The initial reservation wage equals the normalized wage. Favorable tradable expansion can move the economy above the floor; contraction can leave hours unused at the floor. The production comparison sets labor and agent taxes and both support components to zero, keeping fiscal assumptions out of these maps. The general fiscal result remains Proposition \ref{prop:fiscal}.

At the zero-agent reference allocation, the sufficient local bargaining thresholds implied by the main proxies are approximately .145 in $N$ and .659 in $T$ for Brazil, and .174 in $N$ and .451 in $T$ for Mexico. Appendix \ref{app:quantitative} reports their measurement sensitivity.

Both gain and loss regions appear. A low supplier weight and sufficiently cheaper execution can generate a favorable tradable response and consumption gain. With a higher supplier weight, the price paid approaches the human fallback and further foreign cost reductions pass through weakly. Delegating more tasks to foreign agents can then lower income and consumption together. There are also regions in which income falls while consumption rises: favorable input terms of trade compensate consumers for reduced domestic earnings.

Initial-price value added requires an additional measurement choice when an input has no initial transactions. The lower volume comparison in Appendix Figure \ref{fig:volume_reference} values imported execution at its human-parity entry quote. The source files also calculate the upper volume measure obtained with the lowest initial quote consistent with the same bargaining rule, the terminal foreign cost, and nonincreasing foreign costs. This reference-price range changes volume comparisons without changing income, actual consumption, or the allocation.

The country contrast reflects both expenditure composition and the nonlabor-share proxy. Mexico's high tradable nonlabor share makes its modeled output response less elastic than Brazil's under that proxy. This sensitivity is economically central: observed operating surplus need not represent a fixed factor. Appendix \ref{app:quantitative} therefore varies tradable responsiveness separately from the accounting proxy and reports an envelope across source years, classifications, and labor-income treatments.

\subsection{Capacity benchmarks and export-price stress}
The accounting proxy can be replaced by an externally referenced supply response. Conditional on the execution cost $G_T$, our supply equation depends on $p_T/G_T$, implying
\begin{equation}
 \left.\frac{\partial\log Y_T}{\partial\log p_T}\right|_{G_T}
 =-\left.\frac{\partial\log Y_T}{\partial\log G_T}\right|_{p_T}
 =\varepsilon_T^S=\frac{1-\alpha_T}{\alpha_T}.
 \label{eq:supply_transport}
\end{equation}
Within this supply equation, an output-price elasticity also measures the response to a fall in execution costs.

\begin{figure}[htbp]
\centering
\includegraphics[width=\textwidth]{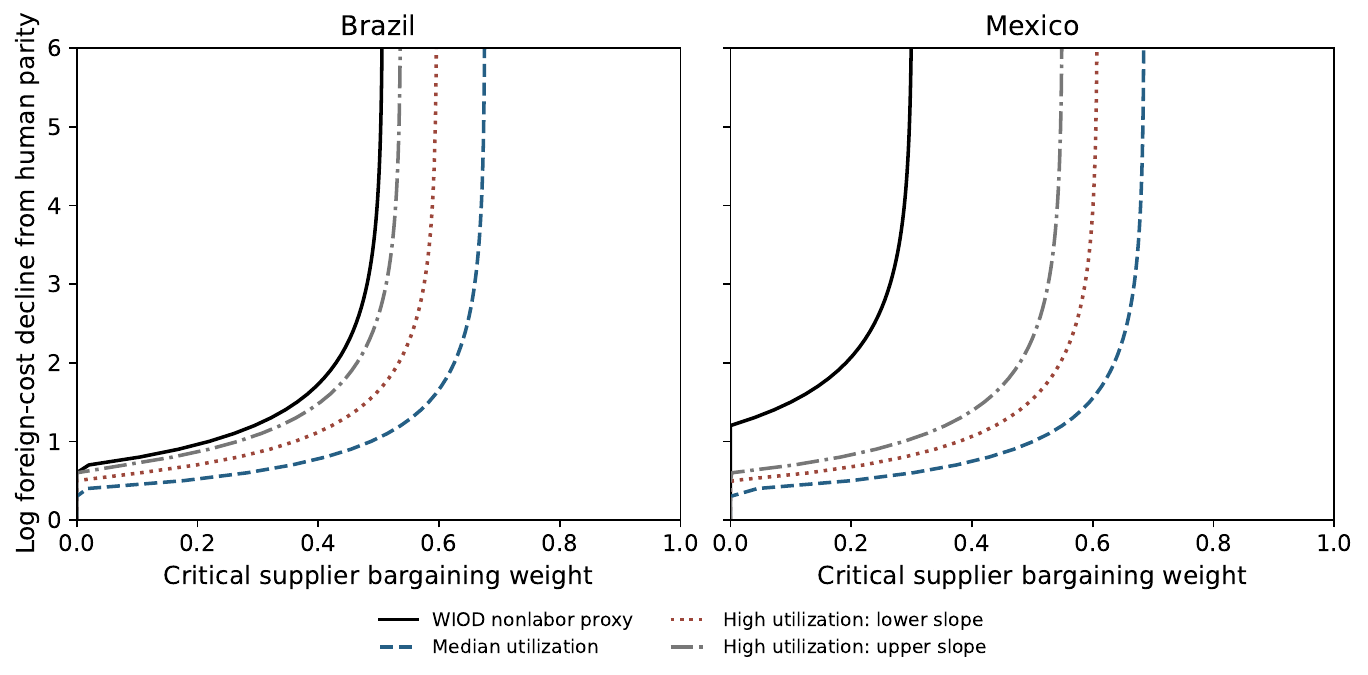}
\caption{Consumption frontiers with external capacity benchmarks. Both task frontiers reach one half. The solid line uses the WIOD nonlabor proxy; other lines conditionally transport the inverse supply slopes .30, .45, and .61 from the March 2020 version of Boehm and Pandalai-Nayar. Consumption falls on the higher-supplier-weight side.}
\label{fig:capacity_benchmark}
\end{figure}

\citet{BoehmPandalaiNayar2022} document capacity-dependent supply in US manufacturing. Their March 2020 working-paper version reports inverse supply slopes near .30 at median utilization and .45--.61 at high utilization \citep[p.~3]{BoehmPandalaiNayar2020}. Figure \ref{fig:capacity_benchmark} transports these three slopes through \eqref{eq:supply_transport}: for inverse slope $s$, set $\alpha_T=s/(1+s)$ and rebase the same initial wage and hours. Country expenditure weights and local-sector parameters are retained. This replaces the tradable compensation-share target. It produces a more responsive tradable sector than the main proxy and enlarges the consumption-gain region, particularly for Mexico.

The source estimates concern US manufacturing, where both markups and marginal costs adjust with capacity utilization. Applying them here assumes the same response to output prices and execution costs in equation \eqref{eq:supply_transport}.

\begin{figure}[htbp]
\centering
\includegraphics[width=\textwidth]{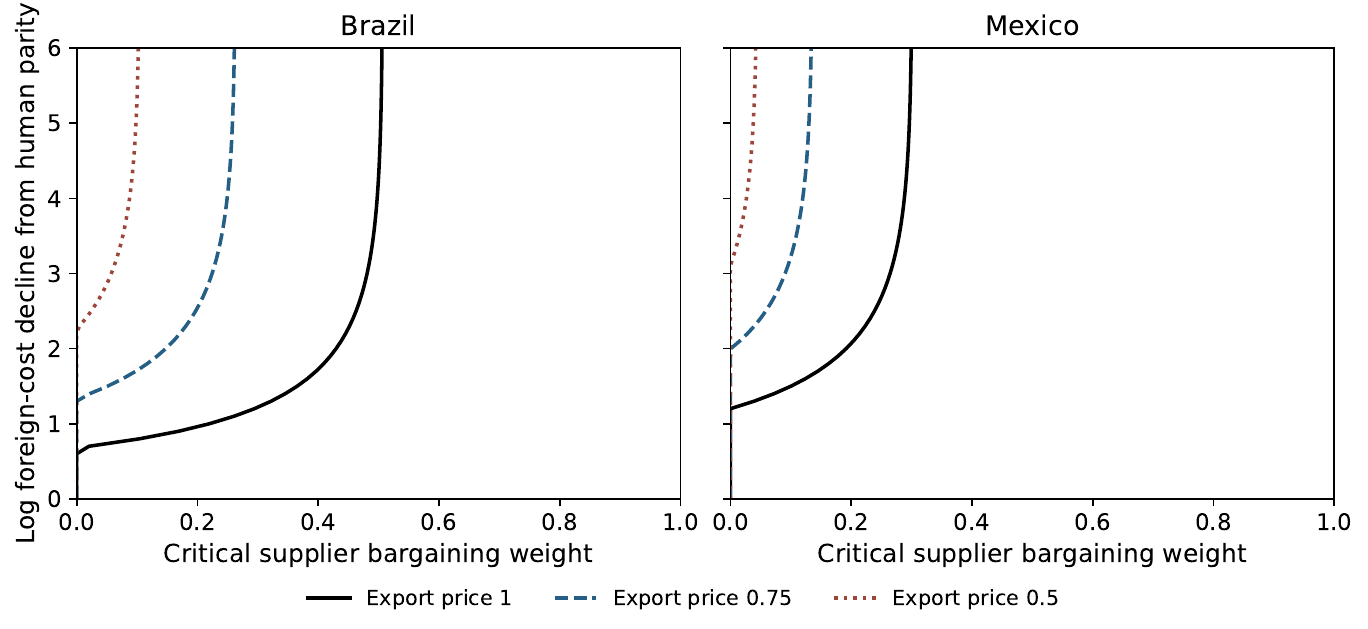}
\caption{Consumption frontiers under an external export-price stress. Both task frontiers reach one half; the tradable price is 1, .75, or .50 relative to its initial value. Reservation earnings and foreign costs use a fixed external unit. Consumption falls on the higher-supplier-weight side. Each allocation solves its labor regime and external account; price movements are exogenous scenarios.}
\label{fig:export_price_stress}
\end{figure}

Figure \ref{fig:export_price_stress} applies Section \ref{sec:export_prices} to the main profiles. It compares export-price ratios of one, three quarters, and one half, keeping $r$, $B_U$, $b$, and each endpoint's foreign marginal costs in the same external unit. We impose these price declines as external stress scenarios. A lower export price shifts the consumption frontier toward lower supplier weights, reducing the set of technology and supplier combinations that deliver a consumption gain. The source files retain the corresponding income, paid-hours, and reference-price volume comparisons.

\subsection{Reservation wages and CES task aggregation}
\begin{figure}[htbp]
\centering
\includegraphics[width=.95\textwidth]{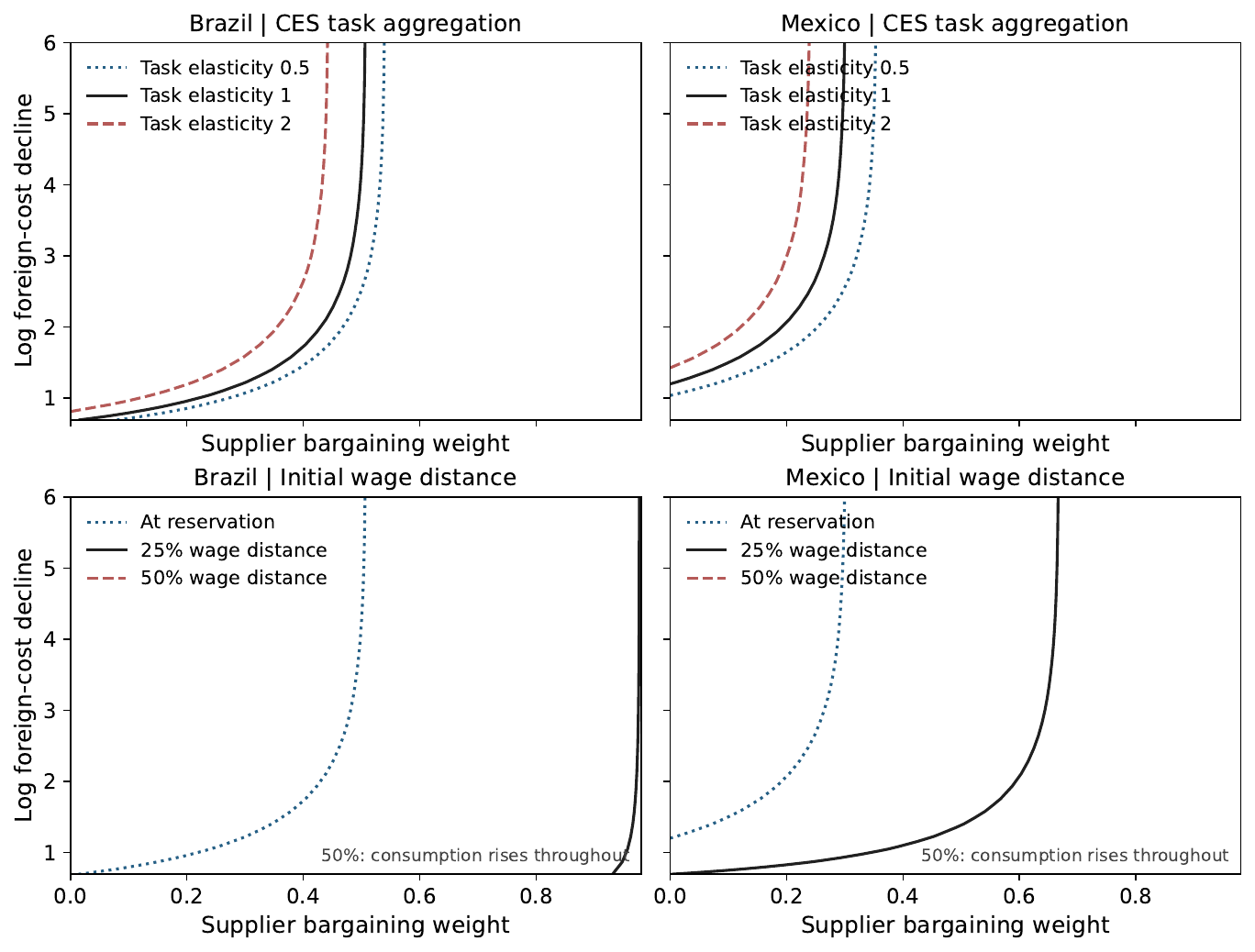}
\caption{Consumption frontiers under alternative task elasticities and reservation wages. Curves mark unchanged consumption, with losses on the higher-supplier-weight side when a crossing exists. The upper panels vary the CES task elasticity at an initially binding reservation wage. The lower panels vary the initial wage distance under Cobb--Douglas tasks. Both frontiers reach one half of tasks; the same initial allocation is matched. The cost domain begins just above $\log 2$ to keep agent execution strictly cheaper than human execution at every displayed reservation wage. Missing curves are identified when consumption rises throughout the displayed range.}
\label{fig:country_robustness}
\end{figure}

A positive reservation remuneration bounds wage adjustment. Its existence leaves open how far the initial wage is above that bound. Figure \ref{fig:country_robustness} matches the same initial wage, paid hours, and zero agent imports, and compares initial wage distances of zero, one quarter, and one half. Foreign costs remain the same across those comparisons. The common displayed domain satisfies strict sourcing at the lowest reservation wage. A larger initial distance permits more wage adjustment before paid hours contract and shifts the consumption-loss frontier. In some comparisons no consumption loss appears in the displayed supplier range. Before the reservation threshold binds, the economy adjusts through wages and paid hours stay constant.

The CES comparisons use task elasticities of one half, one, and two, making the imported expenditure share endogenous as in Appendix \ref{app:CES}. The initial allocation is identical because no task initially uses an agent. Both adverse and favorable regions remain, while their boundaries move.

\subsection{Observable markers and the scenario axes}
Three public indicators help interpret the technological scenarios. The Stanford AI Index reports that the quoted inference price for models reaching a GPT-3.5-level MMLU score fell from \$20 to \$0.07 per million tokens between November 2022 and October 2024 \citep{StanfordAIIndex2025}. These are per-token list prices at a given benchmark performance; the price paid for agent execution can differ.

The ILO's task-based index measures occupational exposure \citep{GmyrekEtAl2025}. It helps select activities for measurement, but the employment share in exposed occupations is a different denominator from $m_j$, the sector's measure of tasks actually delegated to agents. Anthropic's September 2025 report classifies 77 percent of its sampled first-party API conversations as automation patterns \citep{AnthropicEconomicIndex2025}. The unit is a conversation among selected users. Measuring national task coverage requires representative records of delegated work.

Appendix \ref{app:observable_markers} maps the observable objects to the model. Implementation would require sector-specific records of completed tasks, effective execution prices, domestic human alternatives, and the foreign supplier component. Once wages hit the floor, observing $P_j$ and the human fallback cost identifies the saving $\delta_j$. Identifying the supplier share also requires information about foreign marginal cost or bargaining.

A country application would also need participation evidence for the reservation threshold and current expenditure accounts. The source archive retains SEDLAC earnings series as wage references. A fiscal application would additionally need benefit rules, tax coverage, and a tax ceiling.

%% file: tables/country_moments.tex
\begin{tabular}{lrrrrrr}
\toprule
 & $\gamma_T$ & $\gamma_N$ & $\gamma_R$ & $\alpha_T$ & $\alpha_N$ & $\alpha_R$\\
\midrule
Brazil & 0.364 & 0.354 & 0.283 & 0.417 & 0.506 & 0.424\\
Mexico & 0.326 & 0.419 & 0.254 & 0.795 & 0.675 & 0.573\\
\bottomrule
\end{tabular}

%% file: joint_proofs.tex
\subsection{Proof of Proposition \ref{prop:joint} and the price-speed comparisons}
At $t=0$, $\mu_j=\beta_jm_j$ and $\beta_T(1-m_T)+\alpha_T=1-\mu_T$. Substitution into \eqref{eq:sl} gives
\begin{equation}
 s_L+\frac{\alpha_T\Xi}{1-\mu_T}
   =\gamma_T+\beta_N\gamma_N+\beta_R\gamma_R=B,
 \label{eq:Bidentity}
\end{equation}
proving the simplification of $\overline\delta_N$. At the floor and fixed foreign costs, $\dd\log G_j=-\delta_j\dd m_j$, so
\begin{equation}
 \dd\log Y_T=\frac{\beta_T}{\alpha_T}\delta_T\dd m_T.
\end{equation}
Differentiation of the multiplier then gives \eqref{eq:jointE}. From $wL=s_LE$, $w\dd L/E=s_L\dd\log E+\dd s_L$. The production envelope yields
\begin{equation}
 \frac{\dd V^Q}{E}
 =s_L\dd\log E+\dd s_L
   +\frac{\beta_T\Xi}{1-\mu_T}\delta_T\dd m_T
   +\beta_N\gamma_N\delta_N\dd m_N.
\end{equation}
Using \eqref{eq:sl_derivatives}, the $T$ coefficient factors as
\begin{equation}
 \beta_T\left(\frac{s_L}{\alpha_T}+\frac{\Xi}{1-\mu_T}\right)
     \left(\delta_T-\frac{\alpha_T}{1-\mu_T}\right)
   =\mathcal A_T(\delta_T-\overline\delta_T),
\end{equation}
where \eqref{eq:Bidentity} gives $\mathcal A_T=\beta_TB/\alpha_T$. The $N$ coefficient is $\mathcal A_N(\delta_N-\overline\delta_N)$, proving \eqref{eq:jointV}.

The bargaining rule implies $P_j/(\underline w/h_j)=\zeta_j+(1-\zeta_j)c_{Aj}/(\underline w/h_j)$ and hence $0<\delta_j<-\log\zeta_j$. The strict bargaining thresholds therefore make each frontier coefficient negative at every eligible cost level on the feasible floor branch. Income falls by \eqref{eq:jointE}; the wage-bill share also falls, so hours fall. Equation \eqref{eq:realconsumption} gives the real-consumption sign. The ordering follows from
\begin{equation}
 \overline\delta_N-1
 =\frac{\beta_N\gamma_N(1-m_N)+\beta_R\gamma_R}{\Xi}>0,
 \qquad
 1-\overline\delta_T
 =\frac{\beta_T(1-m_T)}{1-\mu_T}>0.
\end{equation}
At equality in a bargaining threshold, the corresponding coefficient remains negative at every strictly positive eligible foreign cost and tends to zero as that cost vanishes. Strict inequality provides a uniform negative margin. These are local statements at equilibrium task shares, not guarantees that all changes in costs preserve the labor regime or fiscal feasibility. The strict signs persist under sufficiently small positive agent taxes by continuity; the displayed closed-form bounds apply at $t=0$.

For the stronger gross-output claim, nontradable expansion alone gives $\dd\log E=-c_N\dd m_N$ and $\dd\log Y_N=\beta_N(\delta_N-c_N)\dd m_N$. Subtracting $c_N$ from $B/\Xi$ gives \eqref{eq:intervalpositive}. The bargaining gap can range over $(0,-\log\zeta_N)$ as foreign cost varies, giving the stated overlap condition. Under simultaneous frontier changes, $\dd\log Y_N=\beta_N(\dd\log E+\delta_N\dd m_N)$.

Allowing foreign price changes adds $-\beta_Tm_T\dd\log P_T/\alpha_T$ to \eqref{eq:jointE}. The resulting hours effect and production envelope give \eqref{eq:joint_costV}. In \eqref{eq:realconsumption}, the foreign bills satisfy $P_Tq_T/E=\mu_T\Xi/(1-\mu_T)$ and $P_Nq_N/E=\gamma_N\mu_N$. Substitution gives \eqref{eq:joint_costQ}. In particular,
\begin{equation}
 \mathcal C_T+\frac{\mu_T\Xi}{1-\mu_T}=m_T\mathcal A_T.
\end{equation}
Finally, $\phi_j=(1-\zeta_j)c_{Aj}/P_j<1-\zeta_j$ because $P_j>c_{Aj}$ in the eligible regime. Combining this bound with $\overline\delta_j-\delta_j>\overline\delta_j+\log\zeta_j$ proves the sufficient speed condition \eqref{eq:uniform_speed}.

%% file: export_price_proofs.tex
\section{Export-price adjustment and reservation-wage crossing}
\label{app:export_prices}
All monetary primitives in this extension use the common external numeraire. In particular, $r,B_U,b,c_{Aj}$ are unchanged when $p_T$ alone changes. The supplier rule remains
$P_j=\zeta_jw/[(1+t)h_j]+(1-\zeta_j)c_{Aj}$.
Define
\begin{equation}
 \theta_j=\frac{\zeta_jw}{(1+t)h_jP_j},\qquad
 \omega_j=1-m_j+m_j\theta_j,\qquad
 \varepsilon_T^S=\frac{\beta_T}{\alpha_T}.
\end{equation}
Then $\dd\log G_j=\omega_j\dd\log w$ for the pure export-price experiment. The task shares, $s_L$, and $\Xi$ are constant. Tradable supply and aggregate labor demand obey
\begin{align}
 Y_T&=K_TA_T^{1/\alpha_T}
       \left(\frac{\beta_Tp_T}{G_T(w)}\right)^{\varepsilon_T^S},\\
 L^d(w;p_T)&=\frac{s_L(1-\mu_T)p_TY_T(w;p_T)}{\Xi w},\\
 \dd\log L^d&=(1+\varepsilon_T^S)\dd\log p_T
                -(1+\varepsilon_T^S\omega_T)\dd\log w.
 \label{eq:price_labor_derivative}
\end{align}
The strictly negative wage derivative of labor demand guarantees a unique labor-market allocation. On the full-employment branch, setting $\dd\log L^d=0$ yields \eqref{eq:price_wage}. Writing its right-hand side as $\psi_T$, the wage-bill identity gives $\dd\log E=\psi_T\dd\log p_T$. Since $C_T=\gamma_TE/p_T$, $Y_N\propto(E/G_N)^{\beta_N}$, and $Y_R\propto(E/w)^{\beta_R}$,
\begin{equation}
 \left.\frac{\partial\log Q}{\partial\log p_T}\right|_{L=\bar L}
 =\gamma_T(\psi_T-1)
    +\beta_N\gamma_N(1-\omega_N)\psi_T\ \ge0.
 \label{eq:price_full_consumption}
\end{equation}
It is strictly positive if agents are used in at least one of $T,N$ with positive foreign cost and $\zeta_j<1$. At zero agent tax, the production envelope for fixed technology gives $\dd V^Q=w\dd L=0$ on this branch. The consumption effect can therefore differ from the local production-volume effect.

At the floor, $w,G_T,G_N$ are fixed in external units. Hence $Y_T\propto p_T^{\varepsilon_T^S}$ and $E,L\propto p_T^{1+\varepsilon_T^S}=p_T^{1/\alpha_T}$. This proves the income and hours derivatives in Section \ref{sec:export_prices}. The same consumption expressions give $\partial\log Q/\partial\log p_T=B/\alpha_T-\gamma_T>0$. With $t=0$, the envelope gives $\dd V^Q/E=s_L\dd\log p_T/\alpha_T$. These volume formulas use reference prices at the equilibrium being differentiated.

The regime boundary has an exact expression. Holding the technology, taxes, foreign costs, and reservation threshold fixed, calculate $L_f^1=L^d(\underline w;p_T=1)$. Then
\begin{equation}
 p_T^*=\left(\frac{\bar L}{L_f^1}\right)^{\alpha_T}.
 \label{eq:export_price_crossing}
\end{equation}
For $p_T<p_T^*$ the floor binds with unused hours; for $p_T>p_T^*$ all potential hours are employed at a higher wage. If the initial price is one and its full-employment wage exceeds the floor, $p_T^*<1$. Fiscal feasibility is a separate condition throughout.

The external identity follows from $p_TX=I$ and $X=Y_T-\gamma_TE/p_T$. It gives \eqref{eq:price_hurdle}. At a common floor, $C_{T1}/C_{T0}=\mathcal E/\varpi$, $Y_{N1}/Y_{N0}=(\mathcal E/g_N)^{\beta_N}$, and $Y_{R1}/Y_{R0}=\mathcal E^{\beta_R}$. Multiplying the consumption ratios proves \eqref{eq:price_consumption_hurdle}. If wages differ between endpoints, its right-hand side acquires the factor $(w_1/w_0)^{-\beta_R\gamma_R}$. The code reports the corresponding wage-conditional consumption hurdle as an ex-post identity.

Finally, at fixed financing and $t=0$,
\begin{equation}
 E\dd\log Q=\dd V^Q+X\dd p_T-\sum_{j=T,N}q_j\dd P_j.
 \label{eq:export_price_envelope}
\end{equation}
To see this, differentiate $E=\sum_jp_jY_j-\sum_jP_jq_j$ and subtract $E\sum_j\gamma_j\dd\log p_j$. Local goods clearing cancels both local price terms, while the tradable price term becomes $(Y_T-C_T)\dd p_T$. The export-price exposure is net exports. Neither gross tradable production nor the accounting multiplier alone measures this terms-of-trade exposure.

The numerical implementation converts all external monetary primitives into tradable units, solves the benchmark equilibrium, and converts monetary outputs back. An independent root of the primitive goods and factor equations checks this conversion. Centered differences check both labor branches, the volume and consumption envelopes, finite compensation, and the crossing price. A consumption-indexed threshold, endogenous global export demand, and an equilibrium world-price response to AI require additional assumptions.

%% file: fiscal_details.tex
\section{Universal and hours-contingent support}
\label{app:fiscal_details}
\label{sec:fiscal}
The support rule $B_U+b(\bar L-L)$ contains a universal allocation ($b=0$), an hours-contingent benefit ($B_U=0$), and mixtures of the two. The population is fixed, so $B_U$ is the aggregate cost of fixed per-resident grants. All monetary parameters remain fixed in the external numeraire. The universal grant is received with or without employment; only the supplement $b$ is withdrawn when an hour is worked. At the maintained $r$, changing either fiscal parameter leaves the real allocation unchanged whenever the budget can be financed.

\begin{proposition}[Fiscal feasibility with two support components]
\label{prop:fiscal}
At the reservation wage with $F=0$, define $\kappa=b/\underline w$, $\rho=\kappa+\tau_L\le1$, $J=I/E$, and $\Gamma=\rho(\alpha_T+t)-t$. Required VAT is
\begin{equation}
 \tau_V^{\rm req}=\frac{B_U+b\bar L}{E}-\rho s_L-tJ.
 \label{eq:VATfloor}
\end{equation}
For fixed support parameters, taxes, and reservation earnings, a frontier change satisfies
\begin{equation}
 \dd\tau_V^{\rm req}
 =-\frac{B_U+b\bar L}{E}\dd\log E+\Gamma\dd J.
 \label{eq:VATderivative}
\end{equation}
If income falls and task frontiers weakly expand, with at least one strict increase, required VAT rises when $\Gamma\ge0$ and either $B_U+b\bar L>0$ or $\Gamma>0$. Public support is feasible under the imposed tax limit exactly when
\begin{equation}
 B_U+b(\bar L-L)-\tau_LwL-tI\le\bar\tau_V E.
 \label{eq:feasibility}
\end{equation}
\end{proposition}

\subsection{Income loss and displacement}
The universal component contributes $-(B_U/E)\dd\log E$ to the VAT change, including when paid hours are unchanged. The hours-contingent component is exposed to both income loss and displacement:
\begin{equation}
 \dd\!\left[\frac{B_U+b(\bar L-L)}{E}\right]
 =-\frac{B_U+b(\bar L-L)}{E}\dd\log E-\frac bE\dd L.
 \label{eq:support_channels}
\end{equation}
At the floor, extending the local-service frontier with foreign costs fixed reduces both income and paid hours. It therefore raises the burden of existing support and adds contingent spending. A comparison between universal and contingent schemes must specify whether initial spending or the transfer to a nonworker is held equal. At matched initial spending, a decline from $L_0$ to $L_1$ adds $b(L_0-L_1)/E_1$ to required VAT under the contingent scheme relative to a fixed universal grant, conditional on the same real allocation.

The exact finite counterpart of \eqref{eq:VATderivative}, when the same wage floor binds at both endpoints, is
\begin{equation}
 \tau_{V1}^{\rm req}-\tau_{V0}^{\rm req}
 =(B_U+b\bar L)\left(\frac1{E_1}-\frac1{E_0}\right)+\Gamma(J_1-J_0).
 \label{eq:VATfinite}
\end{equation}
The frontier derivatives of the import ratio are
\begin{equation}
 \mathcal J_T=\frac{\beta_T\Xi}{(1+t)(1-\mu_T)^2}>0,\qquad
 \mathcal J_N=\frac{\gamma_N\beta_N}{(1+t)(1-\mu_T)}>0,
 \qquad \dd J=\mathcal J_T\dd m_T+\mathcal J_N\dd m_N.
 \label{eq:fiscal_H}
\end{equation}
For $\rho<1$, the direct coefficient $\Gamma$ is positive exactly when
\begin{equation}
 t<\frac{\rho\alpha_T}{1-\rho}.
 \label{eq:fiscal_tax_bound}
\end{equation}
For $\rho=1$, $\Gamma=\alpha_T>0$. At $t=0$, required VAT rises with falling income and frontier expansion whenever support or labor taxes are positive. With no support and no labor or agent taxes, required VAT is zero. A sufficiently large agent tax can make the direct term negative; VAT then falls only if the revenue effect outweighs the income effect in \eqref{eq:VATderivative}. A further fall in negative required VAT increases the surplus rebate.

\subsection{Participation and the negative-income-tax interpretation}
The universal grant is retained upon working, so it creates no withdrawal wedge. The fixed nominal reservation threshold also excludes its income effect on participation. A household-choice model could generate such an effect. At a fixed wage, $y=w\ell$ implies $b(\bar\ell-\ell)=b\bar\ell-(b/w)y$ for individual labor earnings $y$. Combined with the labor tax, net fiscal support is affine in earnings, with slope $-(\tau_L+b/w)$, giving the negative-income-tax interpretation discussed by \citet{Friedman1968}. The primitive here is withdrawal per hour. If wages change, $b/w$ changes; a constant withdrawal rate per unit of income would be a different rule.

\subsection{Proof and proportional tax treatment}
At the floor, $w=r/(1-\tau_L)$ is fixed, so $\kappa$ and $\rho$ are constant. Dividing the fiscal numerator $B_U+b\bar L-(b+\tau_Lw)L-tI$ by $E$ proves \eqref{eq:VATfloor}. The share identity
\begin{equation}
 s_L=\sum_{j=T,N,R}\beta_j\gamma_j-(\alpha_T+t)J
 \label{eq:fiscal_share_identity}
\end{equation}
and the frontier derivatives in \eqref{eq:fiscal_H} prove \eqref{eq:VATderivative}. The coefficient of $J$ is constant along these changes, which also proves \eqref{eq:VATfinite}. Differentiating the support bill divided by income gives \eqref{eq:support_channels}. Substituting the tax ceiling into the public budget proves \eqref{eq:feasibility}. Setting $B_U=0$ recovers the hours-contingent formulation; additionally setting $b=r$ gives $\rho=1$ and $\tau_V^{\rm req}=b\bar L/E-1+s_K$.

Private procurement compares $w/h$ with $(1+t)P$, while net remuneration is $(1-\tau_L)w$. Equalizing their proportional tax treatment gives
\begin{equation}
 1+t=\frac{1}{1-\tau_L},\qquad
 t^{\rm align}=\frac{\tau_L}{1-\tau_L}.
 \label{eq:alignment}
\end{equation}
At the floor, this makes the sourcing comparison depend on $P$ versus $r/h$. The rule is an accounting comparison. Choosing an agent tax also requires provider incidence, capability effects, and a welfare objective. Section \ref{sec:dependence} gives the foreign-price response to $t$.

%% file: renewal_details.tex
\section{Renewal quantities, concessions, and incidence}
\label{app:renewal_details}
A loss of nontradable capability can reduce execution while leaving its nominal bill unchanged. Consider a floor-branch comparison at fixed task shares, tradable-sector capability, and foreign marginal costs. If only $h_N$ falls, then $E$, $L$, and $I_N=\mu_N\gamma_NE$ remain unchanged, while $P_N$ rises and $q_N$ falls. Writing
\begin{equation}
 \theta_N=\frac{\zeta_N(w/h_N)}{(1+t)P_N}\in(0,1),
\end{equation}
the local real-consumption effect is
\begin{equation}
 \dd\log Q=\gamma_N[\ell_N+\beta_Nm_N\theta_N]\dd\log h_N<0.
 \label{eq:renewalQ}
\end{equation}
The foreign supplier's operating surplus $\Pi_N^F=I_N-c_{AN}q_N$ rises by $c_{AN}q_N\dd\log P_N>0$.

The first term in \eqref{eq:renewalQ} is the loss of human execution efficiency; the second is the adverse renewal-price effect. A constant nominal import bill can therefore conceal a loss of effective execution and a greater foreign rent share. The nominal outflow can stay fixed as execution contracts. A requirement to preserve task quantity or service consumption would produce a different financing need.

Introductory concessions also affect the lifecycle account. Let $\mathbf m_0=(m_{T0},m_{N0})$ index initial adoption. If $C^0(\mathbf m_0)$ is a foreign-funded initial concession and $\mathcal M_1(\mathbf m_0)=\sum_jP_{j1}(\mathbf m_0)q_{j1}(\mathbf m_0)$ the later bill, the direct discounted payment is
\begin{equation}
 \mathcal N(\mathbf m_0)=\vartheta\mathcal M_1(\mathbf m_0)-C^0(\mathbf m_0).
 \label{eq:lifecycle}
\end{equation}
Its repricing component is reduced by concessions that respond to anticipated future rents. Complete ex ante competition could offset that transfer. The model does not assume that the initial offer is unanticipated or that all later rent extraction survives lifecycle competition.

An ad valorem agent tax can partly reduce the supplier's net price. Holding $C^H$ and $c_A$ fixed, \eqref{eq:bargain} implies
\begin{equation}
 P(0)-P(t)=\frac{\zeta C^Ht}{1+t},\qquad
 \frac{P(0)-P(t)}{tP(t)}
 =\frac{\zeta C^H}{\zeta C^H+(1-\zeta)(1+t)c_A}.
 \label{eq:taxincidence}
\end{equation}
For interior bargaining and positive foreign cost, that partial supplier share rises with the cost of the domestic alternative. This does not make capability loss desirable: a larger tax share borne by the supplier can coexist with a larger pretax bill and a worse domestic option. The calculation holds the fallback and quantity fixed. General-equilibrium tax incidence also moves wages, execution, and expenditure.

The public-liquidity extension in Appendix \ref{app:liquidity} values the share of a net financing requirement absorbed by reserves, separately from the supplier rent.

%% file: finite_bounds.tex
\section{Finite bargaining bounds at initial reference prices}
\label{app:finitebounds}
For two floor-branch equilibria with fixed foreign costs and human capability and with $t=0$, let $\Delta m_j=m_{j1}-m_{j0}\ge0$. The quoted foreign prices remain fixed. Define the leakage ratio and expenditure ratio by
\begin{equation}
 \rho=\frac{1-\mu_{T1}}{1-\mu_{T0}}\frac{\Xi_0}{\Xi_1},\qquad
 \mathcal E=\rho\exp\left(\frac{\beta_T}{\alpha_T}\delta_T\Delta m_T\right).
 \label{eq:fixedprice_finiteE}
\end{equation}
The external account implies $Y_{T1}-I_1=\gamma_TE_1$, so initial-price value added and real consumption satisfy
\begin{align}
 \frac{V^Q_1}{E_0}
 &=\gamma_T\mathcal E
   +\gamma_N[\mathcal E\exp(\delta_N\Delta m_N)]^{\beta_N}
   +\gamma_R\mathcal E^{\beta_R},
 \label{eq:fixedprice_finiteV}\\
 \frac{Q_1}{Q_0}
 &=\mathcal E^B\exp(\beta_N\gamma_N\delta_N\Delta m_N).
 \label{eq:fixedprice_finiteQ}
\end{align}
The first formula values the final consumption quantities at the initial producer prices. The second applies the Cobb--Douglas quantity weights. These exact comparisons permit finite changes in both sectors.

The right-hand side of \eqref{eq:fixedprice_finiteV} increases with both cost gaps. Substituting their bargaining limits gives
\begin{align}
 \overline{\mathcal E}
 &=\rho\,\zeta_T^{-(\beta_T/\alpha_T)\Delta m_T},\\
 \mathcal F(\zeta_T,\zeta_N)
 &=\gamma_T\overline{\mathcal E}
   +\gamma_N[\overline{\mathcal E}\zeta_N^{-\Delta m_N}]^{\beta_N}
   +\gamma_R\overline{\mathcal E}^{\beta_R}.
 \label{eq:finiteuniform}
\end{align}
Thus $V^Q_1/E_0\le\mathcal F$, with the supremum approached as both positive foreign marginal costs tend to zero. The condition $\mathcal F<1$ guarantees a finite initial-price value-added decline at every eligible cost pair for which the allocations remain on the feasible floor branch. Weighted arithmetic--geometric mean also gives $Q_1/Q_0\le V^Q_1/E_0$, so real consumption falls under this condition.

Each single-sector finite bound is explicit. For expansion confined to $N$, $\mathcal E=\Xi_0/\Xi_1<1$, and value added falls exactly when $\delta_N<\overline\delta_N^{\,F}$, where
\begin{equation}
 \overline\delta_N^{\,F}
 =\frac{1}{\beta_N\Delta m_N}
   \log\left[\frac{1-\gamma_T\mathcal E-\gamma_R\mathcal E^{\beta_R}}
                   {\gamma_N\mathcal E^{\beta_N}}\right].
 \label{eq:finiteNbound}
\end{equation}
For expansion confined to $T$, the value-added comparison is increasing in $\mathcal E$ and equals one at $\mathcal E=1$. Its exact bound is
\begin{equation}
 \overline\delta_T^{\,F}
 =\frac{\alpha_T}{\beta_T\Delta m_T}
   \log\left(\frac{1-\mu_{T0}}{1-\mu_{T1}}\right).
 \label{eq:finiteTbound}
\end{equation}
In each case, $\zeta_j>\exp(-\overline\delta_j^{\,F})$ guarantees the adverse finite volume sign uniformly over eligible costs. The limits as $\Delta m_j\to0$ recover $B/\Xi_0$ and $\alpha_T/(1-\mu_{T0})$, respectively.

Local volume changes use the prices prevailing at each reference equilibrium. Integrating them gives a chained measure; a final basket valued at the original prices can have a different sign. Equations \eqref{eq:fixedprice_finiteV}--\eqref{eq:finiteTbound} give the initial-price comparison directly. For changing foreign prices, the general finite identity \eqref{eq:finiteVQ} applies instead of the fixed-price simplification.

%% file: ces_robustness.tex
\section{CES aggregation across tasks}
\label{app:CES}
This extension keeps the fixed domestic factors, perfect human--agent substitution within eligible tasks, one labor market, nominal reservation threshold, and external closure. It replaces the Cobb--Douglas task composite with
\begin{equation}
 X_j=\left[\int_0^1x_j(i)^{(\epsilon_j-1)/\epsilon_j}\,\dd i\right]^{\epsilon_j/(\epsilon_j-1)},
 \qquad Y_j=A_jK_j^{\alpha_j}X_j^{\beta_j},\quad\epsilon_j>0.
\end{equation}
The continuous limit at $\epsilon_j=1$ is the benchmark. Here $\epsilon_j$ governs substitution across tasks; eligible tasks retain perfect substitution between their human and agent executors.

Write $C^H_j=w/h_j$ and $C^A_j=(1+t)P_j$. The unit cost and the agent share of execution expenditure are
\begin{align}
 G_j&=[(1-m_j)(C^H_j)^{1-\epsilon_j}+m_j(C^A_j)^{1-\epsilon_j}]^{1/(1-\epsilon_j)},\\
 M_j&=\frac{m_j(C^A_j)^{1-\epsilon_j}}
             {(1-m_j)(C^H_j)^{1-\epsilon_j}+m_j(C^A_j)^{1-\epsilon_j}}.
 \label{eq:CESshare}
\end{align}
The shares become $wL_j=\beta_j(1-M_j)\R_j$ and $\mu_j=\beta_jM_j/(1+t)$. Changing only the cost index while retaining $\mu_j=\beta_jm_j/(1+t)$ would violate these expenditure conditions. The aggregate multiplier continues to hold with the endogenous $\mu_j$.

For fixed foreign prices and a binding reservation wage, define $a_j=-\partial\log G_j/\partial m_j>0$, $M'_j=\partial M_j/\partial m_j>0$, and $\widetilde\delta_j=a_j/M'_j$. If $k_j=1-\epsilon_j$, $\delta_j=\log(C^H_j/C^A_j)$, and $D_j=1-m_j+m_j\exp(-k_j\delta_j)$, then
\begin{equation}
 M'_j=\frac{\exp(-k_j\delta_j)}{D_j^2},\qquad
 \widetilde\delta_j=
 \frac{(1-m_j)[\exp(k_j\delta_j)-1]+m_j[1-\exp(-k_j\delta_j)]}{k_j}.
\end{equation}
The limits are $M'_j=1$ and $\widetilde\delta_j=\delta_j$ at $\epsilon_j=1$.

At $t=0$, let $B$, $\mathcal A_j$ retain their definitions in \eqref{eq:joint_coeffs}, and evaluate $\Xi$ and the bounds in \eqref{eq:joint_bounds} using \eqref{eq:CESshare}. The exact local frontier effects are
\begin{align}
 \frac{\dd V^Q}{E}
 &=\sum_{j=T,N}\mathcal A_j
      (\widetilde\delta_j-\overline\delta_j)M'_j\dd m_j,
 \label{eq:CESvolume}\\
 \dd\log E
 &=\frac{\beta_T}{\alpha_T}
       (\widetilde\delta_T-\overline\delta_T)M'_T\dd m_T
       -\frac{\beta_N\gamma_N}{\Xi}M'_N\dd m_N.
 \label{eq:CESincome}
\end{align}
These follow by differentiating the multiplier and labor demand, and applying the production envelope $\dd V^Q=w\dd L+\sum_j\beta_j\R_j a_j\dd m_j$. Thus nontradable frontier expansion still lowers income at the reservation wage for every positive task elasticity. Its volume sign uses the effective saving $\widetilde\delta_N$. The bound ordering $\overline\delta_N>1>\overline\delta_T$ survives. Converting that ordering into supplier-weight thresholds requires the sector-specific mapping from prices to $\widetilde\delta_j$; the benchmark's closed-form thresholds in $\zeta_j$ should not be carried over unchanged.

For every finite pair of task elasticities, an open adverse region exists. As supplier weights approach one, eligible price gaps and $\widetilde\delta_j$ approach zero, while the bounds remain positive. Both frontier contributions in \eqref{eq:CESvolume} are then strictly negative. By continuity, a strict benchmark sign also persists over a neighborhood of $\epsilon_j=1$. At unchanged foreign prices, the local consumption--volume identity \eqref{eq:realconsumption} continues to hold. Above the reservation wage, the full-employment production envelope gives a positive local volume effect from cost-saving frontier expansion.

The benchmark's price-only simplifications require separate care. At a fixed wage,
\begin{equation}
 \frac{\partial M_N}{\partial\log P_N}=(1-\epsilon_N)M_N(1-M_N).
\end{equation}
A nontradable price decline therefore changes the import share when $\epsilon_N\ne1$, and need not leave income or volume value added unchanged. The numerical extension recomputes sourcing expenditure, quantities and prices at every node. It evaluates finite volume changes at the original prices, without imposing the benchmark's local price decomposition.

The source archive implements this separate CES solver and configurable path runner in \path{ces_robustness.py}. The verification script checks the Cobb--Douglas limit, reconstructs primitive task production, solves independent market equations, and compares both sector derivatives with \eqref{eq:CESvolume}--\eqref{eq:CESincome}. Numerical checks span $\epsilon_j\in\{0.5,0.75,1,1.25,1.5,2\}$ with synthetic reference structures in both labor regimes. They recover strict adverse regions and the positive full-employment volume envelope. These are implementation and existence checks, without an empirical task elasticity or a calibrated country result. The configurable runner supplies the corresponding comparisons once country profiles are specified. One comparison holds structural primitives fixed and records the resulting initial import shares. An optional matched comparison holds the initial wage, paid hours and imported expenditure shares fixed by recalibrating eligibility and supply scales at each elasticity. Its paths expand the same fraction of remaining human tasks. This separates changes in initial country structure from changes in task substitution.

The effective-gap bounds are aggregate production and consumption sign conditions under the stated closure. Interpreting them as planner adoption conditions or corrective-tax formulas requires an explicit welfare valuation and policy incidence.

%% file: quantitative_appendix.tex
\section{Measurement and export-responsiveness sensitivity}
\label{app:quantitative}
The replication archive contains original Brazilian and Mexican WIOD national tables, the Socio-Economic Accounts, source documentation, extraction code, industry assignments, unrounded moments, and endpoint results. Country profiles are reconstructed from source cells. No AI supplier weight or reservation wage is estimated from these accounts.

\begin{figure}[htbp]
\centering
\includegraphics[width=\textwidth]{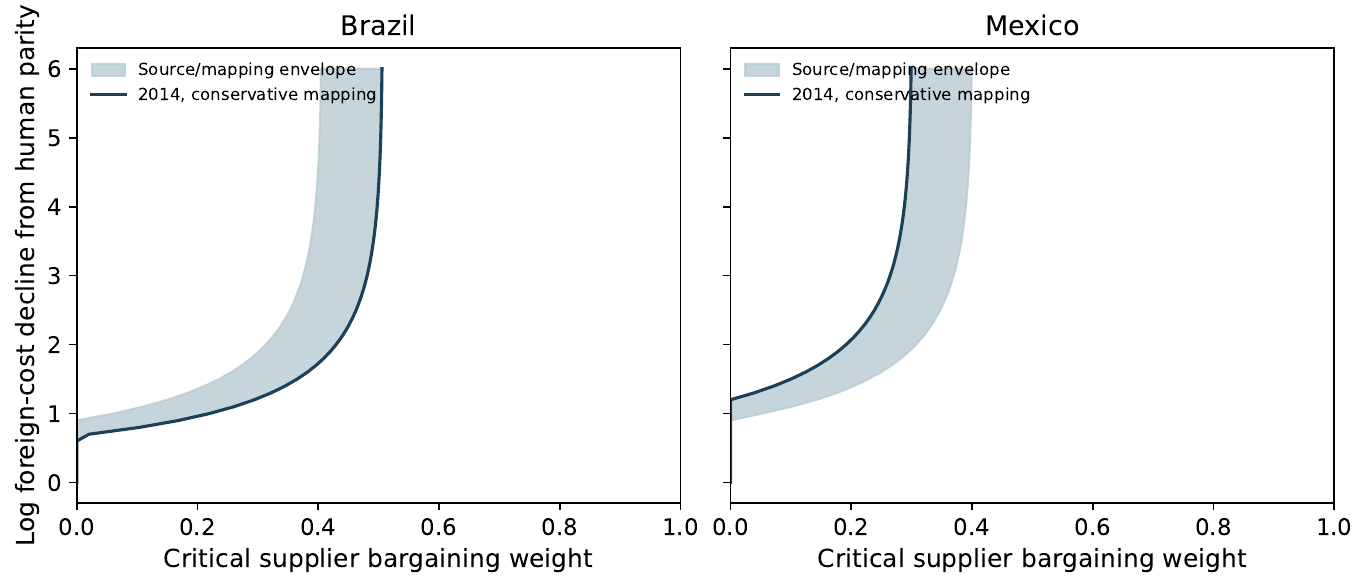}
\caption{Consumption-frontier measurement envelope. Both task frontiers reach one half of tasks. The line uses the main profile; shading spans the disclosed source-year, industry-classification, and labor-compensation choices. For a given profile, consumption falls on the higher-supplier-weight side of its frontier.}
\label{fig:measurement_envelope}
\end{figure}

Figure \ref{fig:measurement_envelope} varies four measurement choices jointly: reference year 2010 or 2014; service export/output cutoff of .05, .10, or .20; administrative-support block assigned to the agentifiable or residual sector; and adjusted labor compensation or employee compensation alone. The last comparison deliberately omits the self-employment correction to expose its consequence, rather than treating the omission as a preferred estimate. Each comparison recomputes both expenditure and factor-income shares. The shaded range is an envelope of model frontiers across these choices, without probabilities or a statistical confidence interpretation.

At the initial zero-agent allocation, the adjusted-labor comparisons give sufficient local nontradable bargaining thresholds between .141 and .280 across countries, years, and classification choices. Including the employee-only sensitivity widens the full range to approximately .141--.295. The tradable thresholds remain higher. These are implications of the model and accounting proxies, not estimates of actual supplier bargaining power. Their sectoral ordering follows analytically from Proposition \ref{prop:joint}.

\begin{figure}[htbp]
\centering
\includegraphics[width=\textwidth]{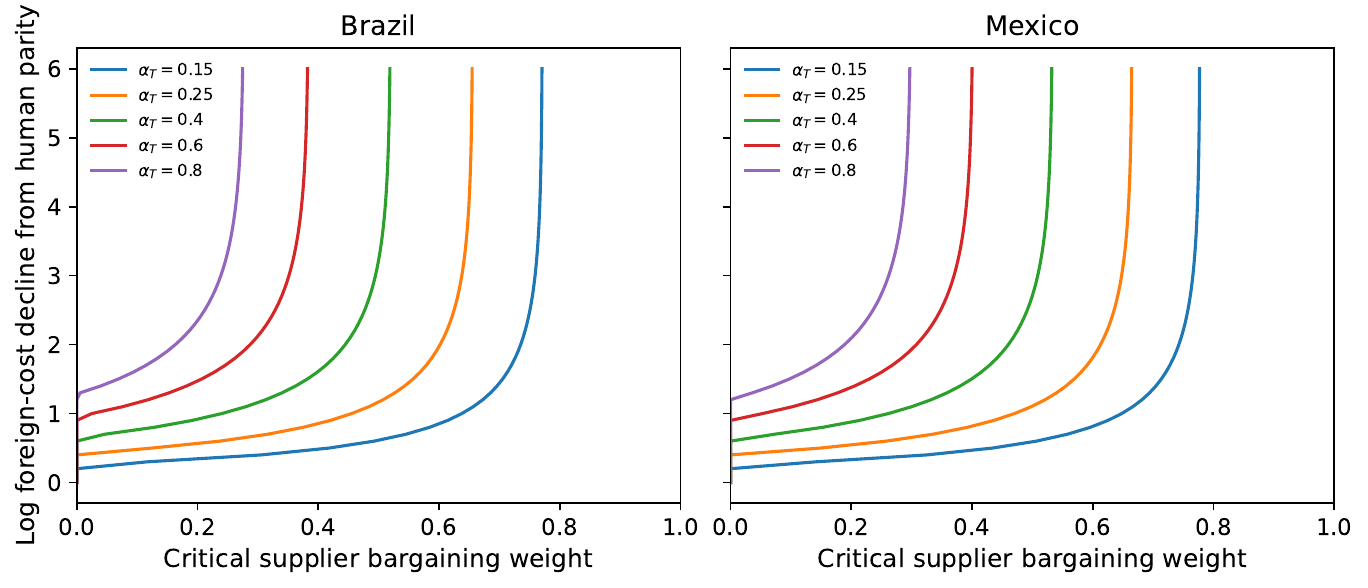}
\caption{Sensitivity to the tradable fixed-factor share. Curves mark unchanged real consumption with both task frontiers at one half of tasks. The displayed $\alpha_T$ values are structural sensitivity inputs; they are not alternative estimates. Consumption falls to the higher-supplier-weight side of each frontier.}
\label{fig:tradable_elasticity}
\end{figure}

Figure \ref{fig:tradable_elasticity} varies $\alpha_T$ directly while retaining the main expenditure weights and local-sector parameters. In this model the tradable execution-cost elasticity is $\beta_T/\alpha_T$. The grid therefore separates a structural export-supply assumption from the empirical nonlabor-share proxy. A more elastic tradable response enlarges the gain region. Country ordering based on a single measured operating-surplus share should not be read as an identified ranking of export adaptability.

\begin{figure}[htbp]
\centering
\includegraphics[width=\textwidth]{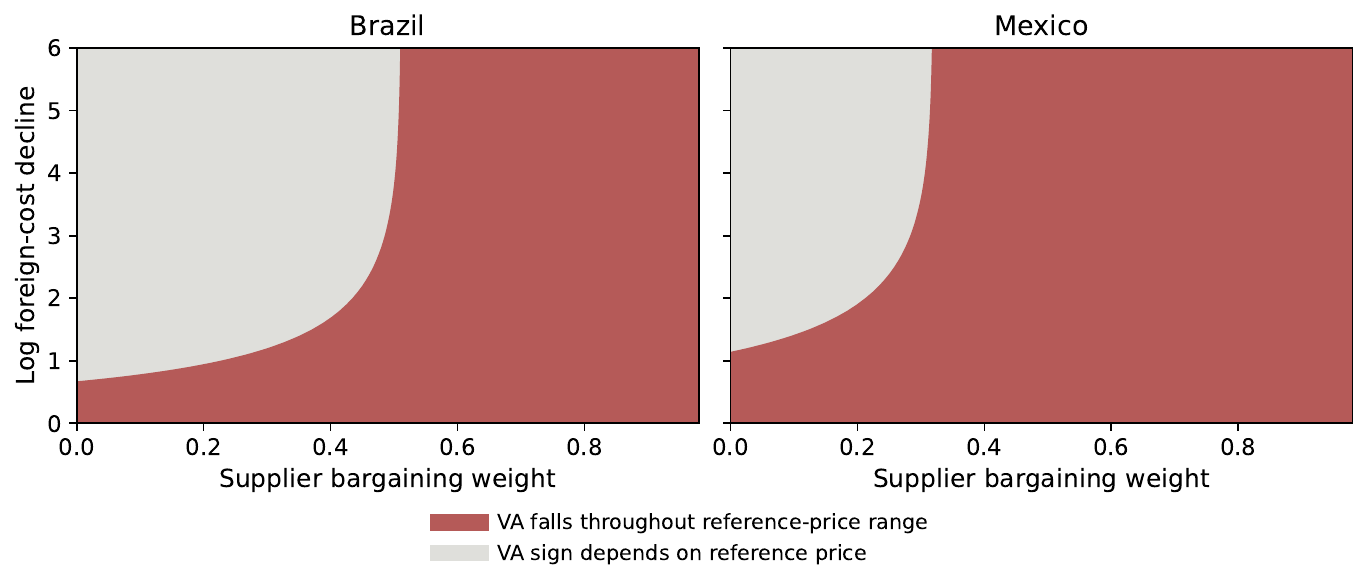}
\caption{Initial-price value-added signs across the admissible input reference-price range. Both frontiers reach one half of tasks. Red identifies a decline throughout the range; gray identifies dependence on the unobserved initial input quote. These comparisons leave actual consumption and income unchanged.}
\label{fig:volume_reference}
\end{figure}

The initial-price input valuation is separately retained in the endpoint files. With $m_0=0$, agent input prices do not affect the initial allocation and cannot be recovered from its transactions. Under a common bargaining weight and nonincreasing foreign costs, an initial quote can lie between $\zeta_j w_0/h_j+(1-\zeta_j)c_{Aj1}$ and $w_0/h_j$. The former gives an upper bound on initial-price value added and the latter the main figure's lower bound. Figure \ref{fig:volume_reference} distinguishes signs that hold throughout that range from signs that depend on the quote. This range concerns a new-input reference-price convention; it leaves the current-income, consumption, import-payment, and labor-regime results unchanged.

%% file: observable_markers.tex
\subsection{Observable objects and model counterparts}
\label{app:observable_markers}
Table \ref{tab:observable_markers} sets out the measurement bridge for future country applications. The cited indicators refer to dated releases; they are not a current-country calibration. No numerical AI input in the maps is fitted to these markers.

\begin{table}[htbp]
\centering\small
\caption{Observable objects and their model counterparts}
\label{tab:observable_markers}
\begin{tabular}{>{\raggedright\arraybackslash}p{.21\textwidth}>{\raggedright\arraybackslash}p{.33\textwidth}>{\raggedright\arraybackslash}p{.37\textwidth}}
\toprule
Model object & Available reference & Additional measurement required\\
\midrule
Foreign marginal cost $c_{Aj}$ & Benchmark-conditioned inference quotes in the AI Index & Supplier resource cost per comparable task; quoted retail prices alone do not identify it.\\[4pt]
Effective saving $\delta_j$ & Buyer invoices and human execution costs & Total agent charge per quality-matched completed task, with verification and integration measured separately.\\[4pt]
Task measure $m_j$ & ILO exposure index; API delegation patterns & Realized task delegation by sector and representative task weights consistent with the production aggregator.\\[4pt]
Foreign share $\mu_j$ & Cross-border supplier invoices & Foreign execution charges divided by sector revenue, excluding domestic integration and separating other imported inputs.\\[4pt]
Tradable response & WIOD moments; external capacity benchmark & Output response to execution-cost changes, conditional on export prices and domestic complements.\\[4pt]
Reservation threshold $r$ & Earnings and participation records & Net reservation earnings for the modeled segment; observed wages alone do not identify the threshold.\\
\bottomrule
\end{tabular}
\end{table}

At $t=0$, the bargaining equation gives
\begin{equation}
 P_j=\zeta_j(w/h_j)+(1-\zeta_j)c_{Aj},\qquad
 \zeta_j=\frac{P_j-c_{Aj}}{w/h_j-c_{Aj}}.
\end{equation}
On the floor branch, $P_j$ and $w/h_j$ identify the cost gap used in the local sign comparison, but leave a continuum of cost--bargaining combinations. Even with a measured saving, the aggregate effect also depends on task coverage, tradable responsiveness, and the labor regime. A fall in a public inference quote therefore cannot locate a country along the marginal-cost axis, and a share of automated conversations cannot locate it along the task frontier.